\documentclass[11pt,a4article]{article}
\pdfoutput=1

\usepackage{jcappub}
\usepackage[latin1]{inputenc}
\usepackage{txfonts}
\usepackage[subnum]{cases}
\usepackage{wasysym}
\usepackage{nicefrac}
\usepackage{multirow}

\usepackage{caption}
\usepackage{subcaption}

\newcommand{\Msun}{\textrm{M}_{\astrosun}}
\newcommand{\lp}{\left(}
\newcommand{\rp}{\right)}
\newcommand{\AU}{\textrm{AU}}

\title{Constraining dark matter sub-structure with the dynamics of astrophysical systems}

\author[1]{Alma X. Gonz\'alez-Morales,}
\author[2]{Octavio Valenzuela,}
\author[3]{Luis A. Aguilar} 
\affiliation[1]{ Instituto de Ciencias Nucleares, UNAM, A.P. 70-543, 04510, Ciudad Universitaria, D.F., Mexico.}
\affiliation[2]{Instituto de Astronomia, UNAM, A.P. 70-264, 04510,  Ciudad Universitaria, D.F., Mexico.}
\affiliation[3]{Instituto de Astronomia, UNAM, A.P. 877, 22860, Ensenada, B.C., Mexico.}

\emailAdd{alma.gonzalez@nucleares.unam.mx, axgonzalez88@gmail.com}

\abstract{
The accuracy of the measurements of some astrophysical dynamical systems allows to constrain the existence of	
incredibly small gravitational perturbations. In particular, the internal Solar System dynamics (planets, Earth-Moon) opens up the
possibility, for the first time, to prove the abundance, mass and size, of dark sub-structures at the Earth vicinity. We find that adopting
the standard dark matter density, its local distribution can be composed by sub-solar mass halos with no currently measurable
dynamical consequences, regardless of the mini-halo fraction. On the other hand, it is possible to exclude the presence of dark
streams with linear mass densities higher than $\lambda_{\rm st}> 10^{-10} \Msun/\AU$ (about the Earth mass spread along the diameter of the SS up
to the Kuiper belt). In addition, we review the dynamics of wide binaries inside the dwarf spheroidal galaxies in the Milky Way. The dynamics of such
kind of binaries seem to be compatible with the presence of a huge fraction of dark sub-structure, thus their existence is not a sharp discriminant of
the dark matter hypothesis as been claimed before. However, there are regimes where the constraints from different astrophysical systems may reveal
the sub-structure mass function cut-off scale.}

\keywords{dark matter theory, power spectrum, particle physics--cosmology connection, dwarf galaxies}

\begin{document}
\maketitle

\newpage

\section{Introduction}

The $\Lambda-\textrm{C$\textrm{DM}$}$ cosmological model assumes the existence of two unknown energy components; a cosmological constant, $\Lambda$,
and the Cold Dark Matter, $\textrm{C$\textrm{DM}$}$. This model is the paradigm on cosmology due to the high precision at which the observations of
the large scale Universe are described (Cosmic Microwave Background radiation~\citep{Komatsu:2008hk}, large surveys of
galaxies~\citep{Tegmark:2006az}, gravitational lensing surveys~\citep{Dye:2007nv}, etc.). Such observations seem to be equally satisfied independently
of the dark matter, ($\textrm{DM}$), nature; although it has to be quite cold (low velocities) \citep{Viel:2007mv}, it can have a small
self-interaction
\citep{Clowe:2006eq}, and at most a weak interaction with the standard model of
particles~\citep{1985NuPhB.253..375S,Jungman:1995df}.~\footnote{Particles like the sterile neutrinos can play a
Warm $\textrm{DM}$ role, which is still viable according to different astrophysical constraints~\citep{Menci:2012kk,AvilaReese:2000hg,Seljak:2006qw}.
The experimental evidence of neutrino oscillations admits its presence, though the cosmological constraints are not conclusive yet,
see~\citep{GonzalezMorales:2011ty,Archidiacono:2011gq,Hamann:2011ge} for detailed analyses.} On the other hand, observations at sub-galactic scales
are sensitive to the particle physics of the $\textrm{DM}$. 

In the current paradigm structure grows from bottom up; the first objects formed at high redshift were some small sub-galactic units that collapsed
from small perturbations in the primordial density field \citep{White:1977jf}. These clumps then merge hierarchically to form the large bound objects
we observe in the local universe today. The properties, mass and scale, of the small structure is, to some extent, determined by the properties of the
specific $\textrm{DM}$ particle physics. Once a primordial power spectrum is fixed, its linear evolution leads to a mass power spectrum with a low end
cut-off, at some wavenumber $k_{\textrm{fs}}$ (mainly determined by the particle free streaming (fs)). Therefore, one can associate a minimum
mass, $m_{\rm min}$, for the smallest gravitationally bound structure that could be formed in the early universe, to the mass contained in a sphere of
radius $r_{\rm fs}\sim \pi /k_{\rm fs}$. For  WIMP's like particles, the mass cut-off is set in the range of $m_{\rm min} \sim 10^{-4}-10^{-12}
\Msun$~ \citep{Green:2003un,Profumo:2006bv} (the specific values depend on the decoupling temperature of the specific model). Notice there are
cases where the mechanism is different, non-thermal particles or resonances can also produce a power spectrum cut-off
\citep{Boyarsky:2008mt,Johnson:2008se}; an example is provided by the axion for which the cut-off is set around
$10^{-18}-10^{-20}\,\Msun$~\citep{Barranco:2010ib,Johnson:2008se}.

The structure formation process makes very difficult to confront the galactic and sub-galactic scales, which are non-linear, with observations.
However, these scales are the ones that contains the more valuable information. To confirm the existence of {\emph{mini-halos}} (sub-halos with a
mass as small as predicted by the power spectrum cut-off), is crucial for the determination of the $\textrm{DM}$ nature, not only because it could
restrict the particle physics of $\textrm{DM}$,  but also because in the case they do exist, they could affect the ongoing direct and indirect
searches for $\textrm{DM}$.

Numerical simulations have been very useful tools to study the non-linear evolution of structure in the larger scales. But, the survival and
distribution of sub-halos (not only the mini-halos but in general), are difficult to study under this scheme,
due to resolution limits and the difficulty in including the physics of baryons, such as supernova feedback~\citep{Read:2004xc}, tidal
striping~\citep{Read:2005zv}, and others~\citep{Zolotov:2012xd,Read:2006fq}. See also
references~\citep{Vogelsberger:2010gd,Klypin:2010qw,Springel:2008cc} for examples of recent techniques to study the smallest reachable scales.
Also, semi-analytic formalisms have been used to study the assembly history of halos, alone or in combination with N-body
simulations,~\citep{Angulo:2009hf,Gao:2005hn}. These formalisms treats the sub-halos as isolated systems, not taking
into account they were part of a bigger structure. Numerical simulations and semi-analytical formalisms seem to be in good
agreement, but on scales much larger than the ones we are interested in this work.

\noindent Also, some of the observational efforts, planed and/or under development, to constrain the mass power spectrum at galactic and
sub-galactic scales are: fits to the fluctuations of gravitational lensing observables, like time delays, the cusp-fold relation
\citep{Moustakas:2009na,Vegetti:2009cz}, and using flexion variance \citep{Bacon:2009aj}; and the identification of ultra faint  galaxies in current
and future surveys \citep{Papastergis:2011xe,Brown:2005xz}. The scale proved by the above observations is suitable to test warm DM
models; however, none of them would actually probe the cut-off scale in the case of cold DM. Perhaps the possible correlation
between the mass cut-off and the direct detection rates, and/or fluxes, of high energy neutrinos from $\textrm{DM}$ annihilation in the Sun
~\citep{Cornell:2012tb}; the correlation between the mass cut-off  and the radius, and surface density, of the galactic halo~\citep{deVega:2010yk};
or from the detection of proper motions from gamma ray point sources \citep{Koushiappas:2006qq,Ando:2008br}, are the observational prospects, up
today, that would probe the smallest scales of the power spectrum and their relation with the $\textrm{DM}$ nature, in the case$\textrm{DM}$is cold
(WIMP-like).

All of the above has motivated several studies about the evolution, specifically the survival, of the {\emph{mini-halos}} under
the dynamical interaction with stars, the galactic potential, or bigger sub-halos~\citep{Diemand:2008in,2010PAN....73..179B,
Angus:2006vp}. Some of these studies reveal that a huge fraction of the mini-halos survive the assembling process~\citep{Diemand:2005vz,
Moore:2005uu}, while others suggest that only a very small fraction of them remain as mini-halos. Nonetheless, the disruption process leads to stream
like structures as remnants~\citep{Zhao:2005mb,Schneider:2010jr}.

In the present work, we go through a complementary methodology that allows to set limits to the mass, and distribution of
mini-halos, or streams, in the Galactic halo. Previous studies have focus on the disruption of the mini-halos (or, the formation of the streams), by
their interaction with stars. Instead, we will focus on what happens to the dynamical system bound to these stars, like planetary systems, binary
companions, etc., due to the dynamical interaction with the sub-structure, ---mini-halos, or already formed streams---. In general, we will study the
dynamical stability of astrophysical systems by considering the rate of injection energy due to the tidal interaction with a family of $\textrm{DM}$
mini-halos and/or streams (perturbers), whose size and abundance can be related to properties of some $\textrm{DM}$ candidates.

Our own planetary system is a good candidate to explore this approach, since the orbital elements are known with outstanding accuracy. For
instance, the Earth-Moon distance has been measured with increasing precision, thanks to the APOLLO-LLR mission~\citep{2012CQGra..29r4005M}, such that
has been used to test gravity theories for a long time. We will estimate the dynamical perturbations that inner Solar System dynamics, Earth-Moon and
Sun-Neptune systems specifically, could suffer by the interaction with a background population of dark sub-structure; and compare with the current
accuracy threshold in order to obtain constraints on the sub-structure properties.
\noindent Other good systems to use within this approach are the wide binaries inside dwarf Milky Way (MW) satellites. The dynamical interaction
between such kind of binaries and $\textrm{DM}$ has been discussed in~\citep{Penarrubia:2010pa} and~\citep{Hernandez:2008iq}, and their survival has
been proposed as a discriminant of the $\textrm{DM}$ hypothesis. Here we review the dynamics of this systems under the dynamical interaction with
mini-halo mass, and/or streams, and study the effect of cumulative encounters.

This work is structured as follows: first, we present the basic equations to compute the energy perturbation on an astrophysical system due to a
single encounter with a mini-halo like perturber (section~\ref{sec:dynamicpert_single}), and with a stream like perturber
(section~\ref{sec:dynamicpert_single_st}). Then we describe the treatment of cumulative encounters by means of Monte-Carlo experiments in section~
\ref{sec:dynamicpert_multi}. In section~\ref{sec:SSc}, and~\ref{sec:Binc}, we use the described method to study the dynamical perturbations to the
inner Solar System, and to the open binaries, respectively. In section~\ref{sec:initialminihalomass} we outline a possible connection between the
linear mass density of streams, with the mass of the initial mini-halo. Finally, we discuss our results in section~\ref{sec:conclusions}. 

\section{Dynamic perturbations on astrophysical systems due to dark sub-structure}
The dynamics of some astrophysical systems is known with great accuracy, such as our own Solar System, that any dynamical perturbation that
could affect it could be constrained. Including the presence of dark sub-structure. These systems would experience an almost negligible effect if
$\textrm{DM}$
were smoothly distributed over the dark halo, however, if $\textrm{DM}$ were mostly in form of mini-halos, or streams, these act as perturbers to the
astrophysical system (the target). The effect of one single encounter could be small, but the cumulative effect of all the halo sub-structure could
play a significant role to the target dynamics. Below, we work out a formalism that allow us to quantify this effect for mini-halos and streams acting
as perturbers. We will work with binary systems as targets, but the reader should keep in mind that all of what is presented here could be applied to
any astrophysical system, just by describing it in the center of mass reference frame.

\subsection{Single encounters with mini-halos}
\label{sec:dynamicpert_single}
If mini-halos has not suffered important disruption, they can be treated as spherical perturbers. For encounters that occur at distances much larger
than the characteristic size of the target, and when the effective interaction time is a lot shorter than the internal dynamical time of the
target,-- say the orbital period of a planet--, the encounter is said to be in the distant and impulsive regime. Within this regime, the average
energy perturbation, per unit mass, to an extended system due to one single encounter is:
\begin{equation}
\langle\Delta E\rangle_{\rm mh}=\dfrac{7\,G^2\,m_{\rm mh}^2}{3\,v_{0}^2\,p^4} a^2 U(p/r_h).
\label{eq:deltaEpoint}
\end{equation}
where, the impact parameter $p$ is the shortest distance from the target center to the mini-halo, $m_{\rm mh}$ is the mass of the
mini-halo, $v_{0}$ is the relative encounter velocity, $a$ is a characteristic size of the target 
(e.g. the mean particle separation), and $r_h$ the characteristic radius of the perturber~\citep{2008gady.book.....B}. The $U$ function is a
dimensionless correction factor that takes into account finite-size effects of the perturber~(eq. (3) in~\citep{1985ApJ...295..374A}); it depends on
the density profile of the perturber and takes values between $0$ and $1$. As the impact parameter becomes larger than the characteristic
size of the perturber, the $U$ function tends to one and the point mass approximation becomes valid. Most of the encounters in our study falls within
this approximation, for the rest we use the $U$ function given in appendix~\ref{sec:appendix1}. The fractional energy perturbation to the
target is the average energy perturbation normalized to the binding energy ($|E_{\rm b}|= G M_c/(2 a)$, with $M_c$ the mass of the central body in
the target), given by:
\begin{equation}
\dfrac{\langle\Delta E\rangle_{\rm mh}}{\left|E_b\right|}=\dfrac{14\,G\,m_{\rm mh}^2}{3\,M_c\,v_{0}^2\, p^4} a^3.
\label{eq:deltaEmini}
\end{equation}
If it happens that the distant approximation does not hold, then we implement the correction described in appendix~\ref{sec:appendix2}.

\subsection{Single encounters with Streams}
\label{sec:dynamicpert_single_st}
Streams are former $\textrm{$\textrm{DM}$}$ mini-halos (sub-halos in general), that have been already disrupted by tidal forces. The
interaction of streams with a bound system of finite size can be seen as the gravitational interaction of it with a structure of cylindrical symmetry.
The
simplest stream one can model is 1-dimensional (1D), i.e. a line of longitude $L$ and linear mass density $\lambda_{\mathrm{st}}$. Nevertheless
more
realistic models include: the finite cross section stream with constant density (CD); and the finite cross section with a central
core (Core) or with a central cusp (Cusp). In equations \eqref{eq:streams} we present the equations we will use to compute the
fractional energy perturbation to the target (normalized to its binding energy), for the different stream models.

\begin{subnumcases}{\dfrac{\langle\Delta E\rangle_{\rm st}}{|E_b|}=\dfrac{4\,G\, \pi^2 \lambda_{\rm st}^2}{M_c\, v^2_{0}\, {\sin\lp
\theta\rp}^2\,p^2 } a^3 \label{eq:streams}}
 \varmathbb{T}\lp\alpha, \psi\rp & (1D)\\
 \label{eq:1d}
 \varmathbb{T}\lp\alpha, \psi\rp \,{\varmathbb{B}^2\lp R_0/p\rp} &  (CD)\\
 \label{eq:CD}
  \varmathbb{C}\lp R_0/p,\alpha,\beta,\psi\rp  & (Core) \\ 
  \label{eq:core}
 \varmathbb{D}\lp R_0/p,\alpha,\beta,\psi\rp & (Cusp). 
 \label{eq:cusp}
\end{subnumcases}
The functions $\varmathbb{T}$, $\varmathbb{B}$, $\varmathbb{C}$ and $\varmathbb{D}$ are dimensionless and contains information about the geometry of
the encounter, (defined by the angles $\theta, \alpha, \beta$ and $\psi$), and about the characteristic radius of the stream, $R_0$. They are the
equivalent of the $U$-function. All of the equations for the finite cross-section streams converge to that of the 1-dimensional stream for large
impact parameters. The reader is referred to appendix \ref{sec:appendix3} for a derivation of equations \eqref{eq:streams} and for a wider explanation
of the different parameters.

\subsection{Cumulative Effect of Encounters: Monte-Carlo simulations of the encounters}
\label{sec:dynamicpert_multi}
Regardless of whether the interaction is with mini-halos or with streams, the fractional energy of the target system will change
as a result of repeated encounters. Here we can distinguish 2 contrasting regimes: in the first, the "diffusive" regime, the
binding energy slowly, but steadily changes as the effect of small individual encounter perturbations accumulates. In the second,
the "catastrophic" regime, a few encounters are enough to unbind the target. Our general approach to study the cumulative effect of encounters
consists of Monte-Carlo experiments, which automatically accounts for the catastrophic and diffusive regimes. These are performed as follows:
\begin{enumerate}
 \item First, we construct individual histories of encounters. 
\begin{enumerate}
 \item We start bay drawing random values for the encounter parameters, (impact parameter, encounter velocity and orientation of the
target, etc.), following their corresponding distribution functions.
\item  We use these parameters to assess the effect of one single encounter trough the fractional energy given by
equation \eqref{eq:deltaEmini} for mini-halos, or equations \eqref{eq:streams} if the perturber is a stream.
\item Now we update the mean size of the target according to the  fractional energy injection just obtained. The relation between fractional energy
input, and the changes in the mean separation of the target is $\left|\Delta E/E\right|=\left| \Delta a / a \right|$ (from the definition, and the
derivative, of the binding energy). 
\item Then we increment the experiment timer by approximately the time the perturber takes to traverse a sphere of radius equal to the impact
parameter t, with a encounter velocity $v_0$: $\Delta t_{\rm enc}=2 p/v_0$.
\item At last we repeat steps (a) to (d) until the time limit (e.g. the life time of the target), is reached, or until the fractional energy reaches
the unity (i.e. the target gets unbound).
\end{enumerate}
\item The above will give us the total fractional perturbation energy to the target, due to one single history of cumulative encounters
between our
target and a given realization of the perturbers distribution. The second part of our Monte-Carlo consist of creating an ensemble of such histories,
(i.e. to repeat the above procedure a number of times $N$), so we get to know the mean value of the total injection energy that
a population of sub-structure produces on the dynamics of a specific target.
\end{enumerate}
\subsection*{Distribution of encounter parameters.}
Let us describe now the distribution functions we use to sample the different parameters. For encounters with a mini-halos, the impact
parameter will be sampled from a nearest-neighbor distribution, (random, Poisson distribution), modulated by the $\textrm{$\textrm{DM}$}$ local
density ($\rho_{\rm dm}$), the mass of the perturbers ($m_{\rm mh}$) and the fraction ($\rm f_{\rm s}$) of $\textrm{$\textrm{DM}$}$ density that is in
form of mini-halos:
\begin{equation}
 f \lp p \rp\,dp = 4 \pi\,\nu \, p^2 \exp^{\frac{- 4 \,\pi\,\nu p^3}{3}}, \;  \mbox{with} \; \nu =
\dfrac{\rm f_{\rm s}\,\rho_{ \rm dm}}{m_{\rm mh}}.
\label{eq:fp}
\end{equation}
On the other hand, for an encounter with a stream we sample the impact parameter from the probability distribution function shown in
Figure \ref{fig:PDF}. We can not use Eq.\eqref{eq:fp}, now modulated now by the linear density of streams, because the final impact
parameter distribution does depend on the sample volume used to generate it, and there is no way to set it a priori. Instead, we adopted a more basic
procedure that does not depend on any sample volume, it is described in appendix \ref{sec:appendix4}. \\

For both types of encounters, with mini-halos or streams, we sample the relative encounter velocity from a Maxwell-Boltzmann
distribution function. We are considering that the distribution of sub-structure is relaxed and well described by a classic distribution
\citep{Schneider:2010jr}. Finally, the encounter geometry parameters ($\theta$, $\alpha$, etc.) are sampled from uniform probability distributions.

\begin{figure}[t!]

        \begin{subfigure}{0.5\textwidth}
                \centering
                \includegraphics[width=\textwidth]{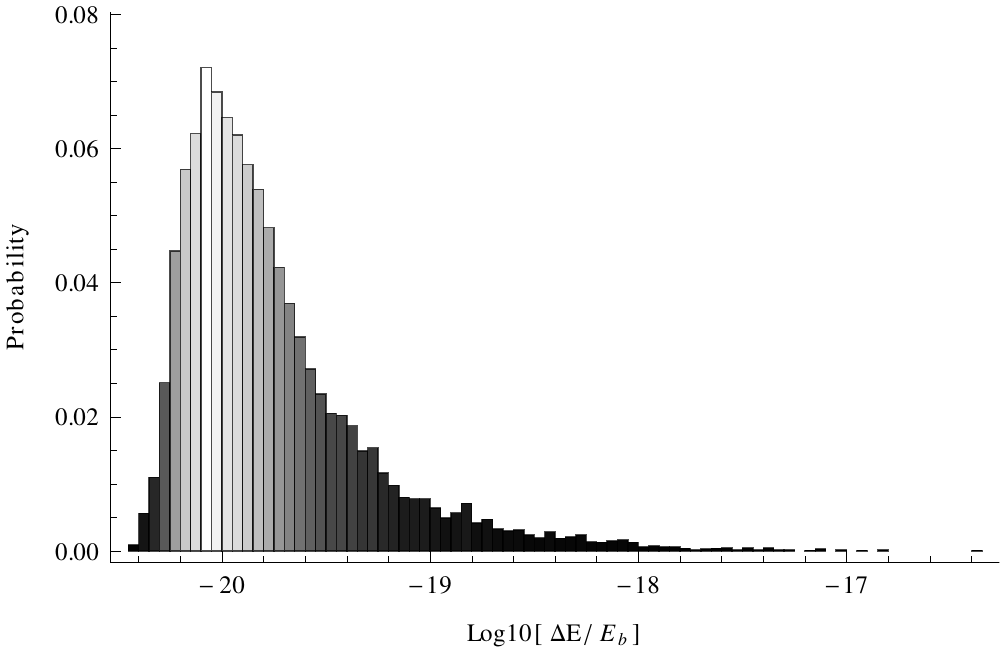}
                \caption{}
                \label{fig:eamoon_example1}
        \end{subfigure}
        \begin{subfigure}{0.5\textwidth}
        \centering
	\includegraphics[width=\textwidth]{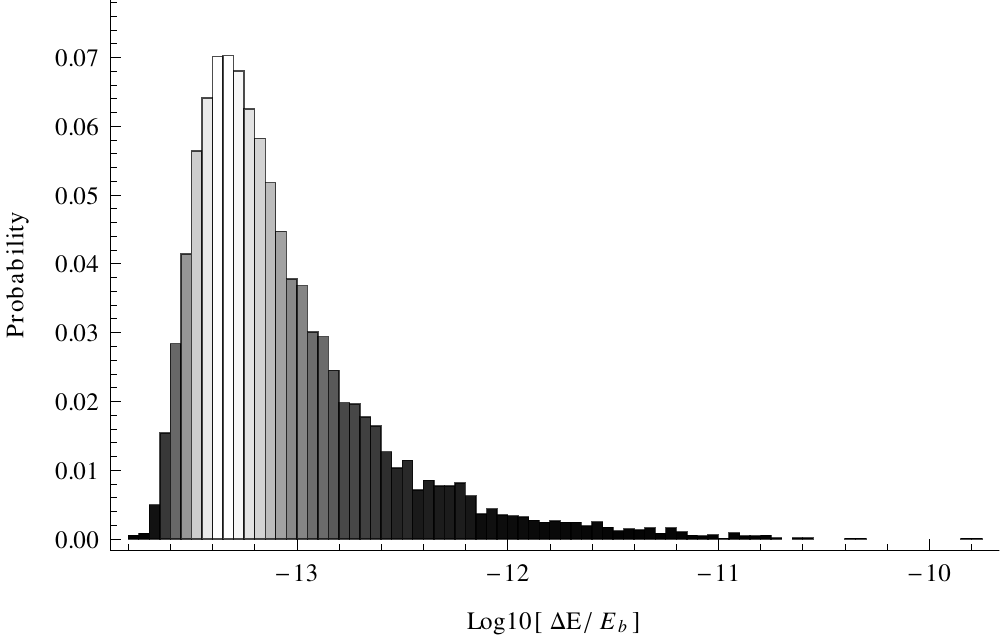}
	\caption{}
	\label{fig:eamoon_example2}
	\end{subfigure}
	\caption{Resultant probability distribution of fractional energy imprinted to the Earth-Moon system due to
	cumulative encounters with: (a) mini-halos of mass $m_{\rm mh}=10^{-6} \Msun/\AU$, and (b) streams of linear density $\lambda_{\rm st}
	=10^{-6} \Msun /\AU$. The ensemble consists of about $10,000$ histories of the target evolution, under the interaction with
the corresponding perturber; for about $4\,\textrm{Gyr}$ each history.}
\label{fig:eamoon_example}
\end{figure}

\section{Application to astrophysical systems}
If a population of $\textrm{$\textrm{DM}$}$ mini-halos and/or streams is pervasive through the Galaxy, as some variants of $\textrm{DM}$ constituents
suggest, it
is likely that their presence could be inferred indirectly from their cumulative dynamical effect on some astrophysical systems,
which could be very sensitive to external perturbations. We now explore the consequences for two such systems.
\subsection{Solar System: Earth-Moon and Neptune orbits}
\label{sec:SSc}
The Solar System dynamics is known with high accuracy. For instance, the uncertainty in the planet semi-major axis is of the
order of $10^{-9} \AU$ for Jupiter and $10^{-6} \AU$ for Neptune~\citep{Pitjeva:2011ts} while for
the Earth-Moon system, it is currently determined with a precision of $\Delta a/a \sim 10^{-13}$~\citep{2012CQGra..29r4005M}.
This precision has made the Solar System to be extensively used to test general relativity, fundamental physics and
alternative theories of gravity (see~\citep[and many others]{Williams:2012nc,Iorio:2012pv} for recent examples), and it is 
particularly attractive to constraint the amount of $\textrm{$\textrm{DM}$}$ that may be present in the Solar
System~\citep{Iorio:2010gh,2010IJTP...49.2506S,2009EAS....36..127J,Frere:2007pi}. It is interesting as well to constrain
the presence of dark sub-structure in the solar neighborhood. 

As the Solar System moves within the Galactic halo, the orbits of planets in this system will experience almost negligible
perturbations if $\textrm{DM}$ is smoothly distributed, contrary to the case if it is distributed in mini-halos or streams. In the later
case, the planet orbits will undergo an energy perturbation due to cumulative encounters with these structures, causing them to
undergo secular evolution, and even to get unbound. We will compare the predicted dynamical effect, changes in the semi-major axis of planet orbits,
principally, with the current accuracy threshold to determine upper limits on the $\textrm{DM}$ sub-structure that
could be present. 

To start with, we choose the Sun-Neptune and  Earth-Moon relative orbits as targets. Both have very stringent constraints in the
determination of their semi-major axis, and they are also suitable to work within the impulsive encounter approximation. Assuming that the encounter
velocity is roughly $200 {\rm km/s}$, (the Sun goes around the galactic center at about $220 {\rm km/s}$ and the dark halo presumably has very
little rotation), means that a perturber crosses Neptune's orbit in just $1.4\,$years, that is many times shorter than Neptune's orbital period of
$164.8 \,$years; we are definitively in the impulsive regime, a similar argument applies for the Earth-Moon System. When considering cumulative
encounters, we assume the encounter relative velocity magnitude to follow a Maxwell-Boltzmann distribution with a mean of $220\,{\rm km/s}$, as
it is usual. Finally, we assume a standard local $\textrm{DM}$ density value, $\rho_{\rm dm}=10^{-18} \Msun\, \AU^{-3}$
($0.35\,\textrm{GeV}/\textrm{cm}^3$)~\citep{Iocco:2011jz,Salucci:2010qr,deBoer:2010eh,Catena:2009mf,Widrow:2005bt} (unless otherwise is specified).

In Figure~(\ref{fig:eamoon_example})  we show a couple of examples of the resultant distribution for the total fractional energy, $\Delta E/E_b$,
obtained in our Monte-Carlo experiments ensembles. Each of the runs in the ensemble corresponds to a history of encounters between the
Earth-Moon system and a population of mini-halos of $m_{\rm mh}=10^{-6} \Msun/{\AU}$ (Figure~( \ref{fig:eamoon_example1})), or streams of
$\lambda_{\rm st}= 10^{-6} \Msun/{\AU}$ (Figure~(\ref{fig:eamoon_example2})). The evolution time of the system in each history is for about 4 Gyr,
approximately the age of the Earth-Moon system, and the time that Neptune has been in its present orbit~\citep{Perryman:2011gz,Crida:2009md}. The
ensembles consists of about $10000$ histories of the target evolution. For both systems the resultant distribution is clearly non-symmetric, then we
will further report the results in terms of the median value and the interquartile range. 

\begin{figure}[t!]
       \begin{subfigure}{0.5\textwidth}
              \centering
             \includegraphics[width=\textwidth]{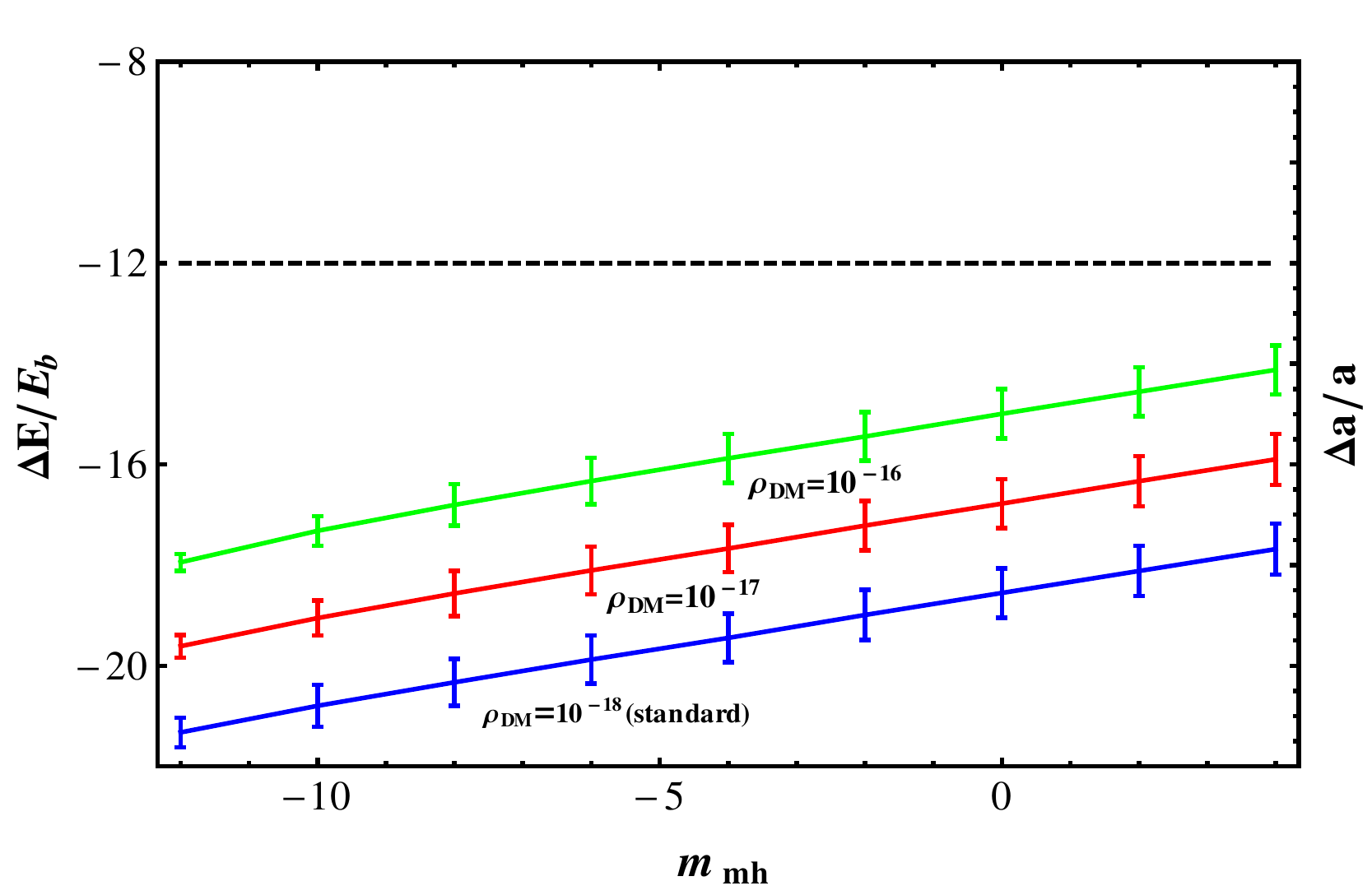}
            \caption{Earth-Moon}
           \label{fig:eamoon_point}
        \end{subfigure}%
        \,
        \begin{subfigure}{0.5\textwidth}
                \centering
               \includegraphics[width=\textwidth]{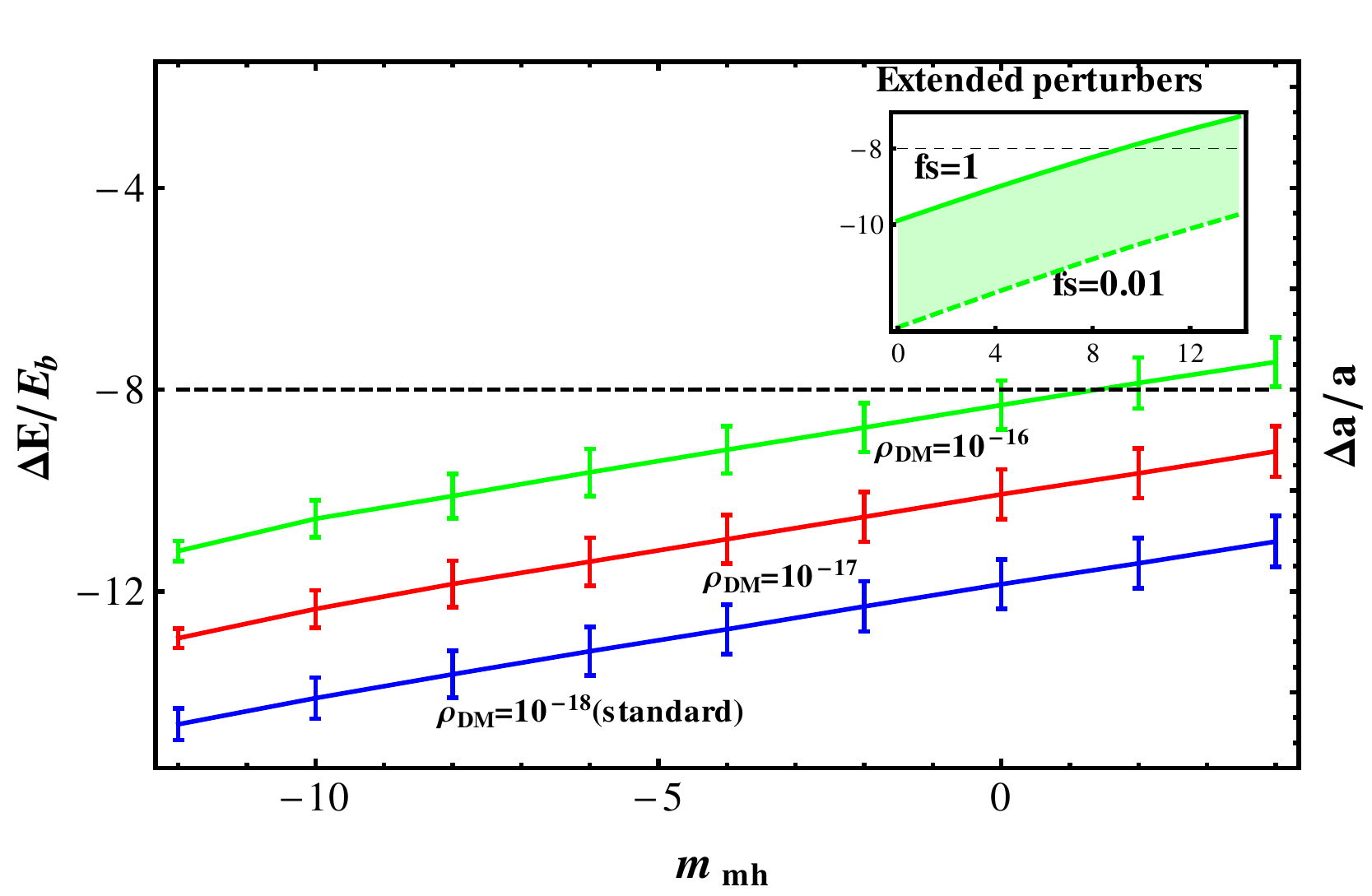}
                \caption{Neptune}	
                \label{fig:neptune_point}
        \end{subfigure}
       \caption{Constraints to the mini-halo mass due to encounters with the (a) Earth-Moon and (b) Neptune System. The median energy
perturbation to the orbit, due to interactions with mini-halos as a function of their mass, compared to the experimental uncertainty in the semi-major
axis of the corresponding orbit\citep{Pitjeva:2011ts,2012CQGra..29r4005M} (dashed line). Three different values of the local $\textrm{DM}$ density,
$\rho_{\rm DM} (\Msun/\textrm{AU}^3)$, are shown for comparison.  The bars correspond to the interquartile range for the Montecarlo realization. The
sub-panel shows the effect of include the correction for extended perturbers, and that of varying the fraction of sub-structure.}

\end{figure}

First, lets take a look at the results for encounters with mini-halos. In Figures~ (\ref{fig:eamoon_point}), and~(\ref{fig:neptune_point}), we
summarize
them, for the Earth-Moon and Neptune-sun systems, respectively. The perturber mass is in the range $10^{-12}<m_{\rm mh}<10^4$. We can
see that none of both systems are sensitive enough to the presence of this kind of sub-structure, neither at the standard local $\textrm{DM}$ density
(blue line in the figures), nor at a greater values (red and green lines). Except for the most massive perturbers, more than $10^2
\Msun$, and in the denser case, $\rho_{\rm dm}>10^{-18} \Msun$. This would be the maximum effect that $\textrm{DM}$ mini-halos can produce, since we
have assumed the fraction of mass in mini-halos equal to one and treated them as point masses. It is interesting to note that even with this extreme
approximations the dynamical effect is not observable. More reliable models for mini-halos are expected to be extended and
diffuse \citep{Diemand:2005vz}. On the other hand, the fraction of substructure depends on the slope of the substructure mass function, $n$, which up
to now it is only accessible by the extrapolation of it from larger scales \citep{Springel:2008cc}.  The case of  $n=-2$ would suggest that there is
no
smooth halo at all, and that all the mass would be contained in sub-halos. Although, current results suggest a slope of $n\sim -1.9$, implying  the
presence of a smooth halo component and a substructure fraction below the unity, this result may be verified by future simulations."
To consider extended mini-halos, and smaller fractions of sub-structure,  would result in lower values to the injection energy, (see
appendix \ref{sec:appendix1}, and the sub-panel in Figure~(\ref{fig:neptune_point})). 

\begin{figure}[t!]
       \begin{subfigure}{0.48\textwidth}
              \centering
               \includegraphics[width=\textwidth]{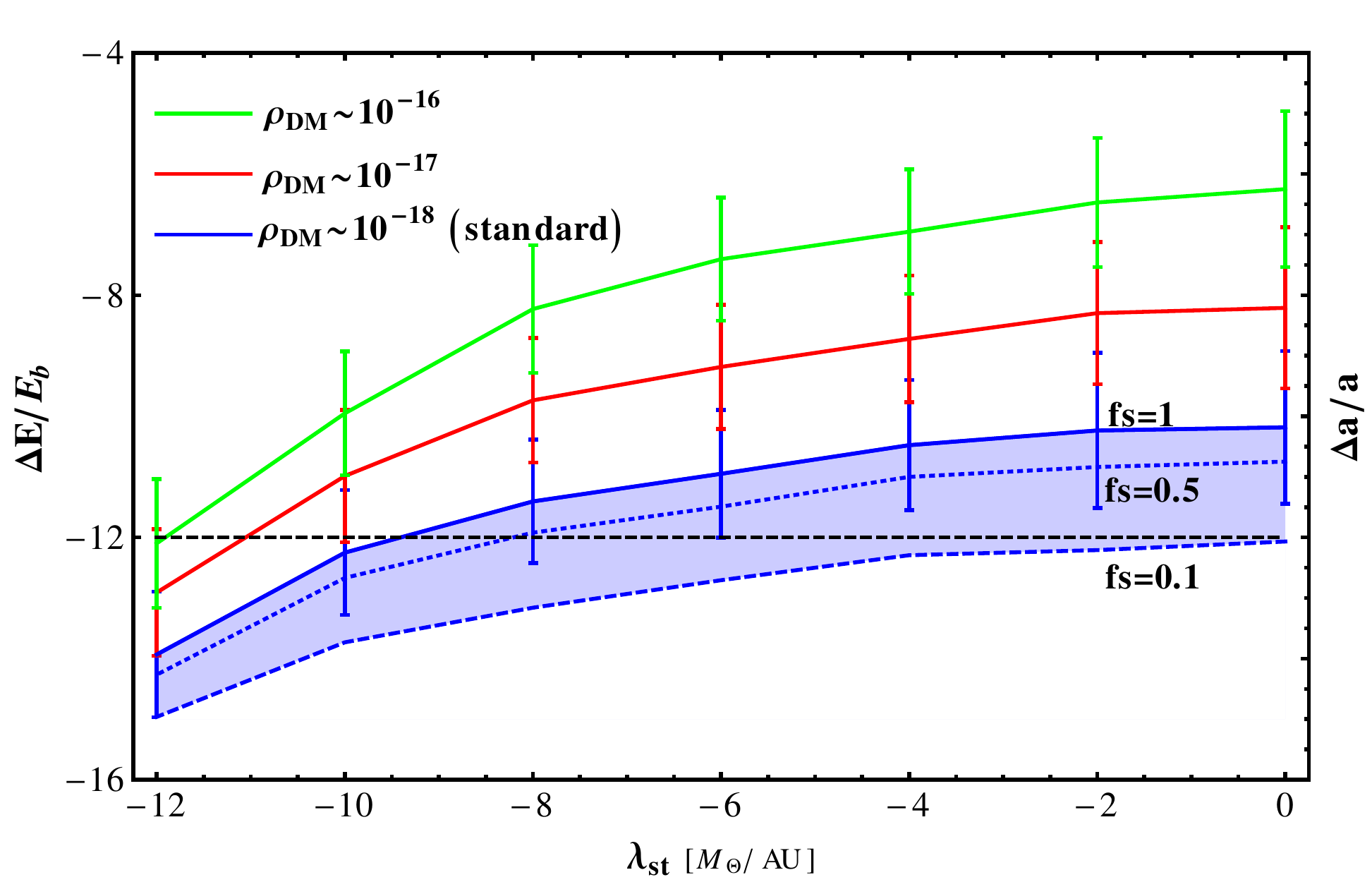}
                \caption{Earth-Moon}
                \label{fig:eamoon_stream}
       \end{subfigure}%
      \,
       \begin{subfigure}{0.48\textwidth}
               \centering
                \includegraphics[width=\textwidth]{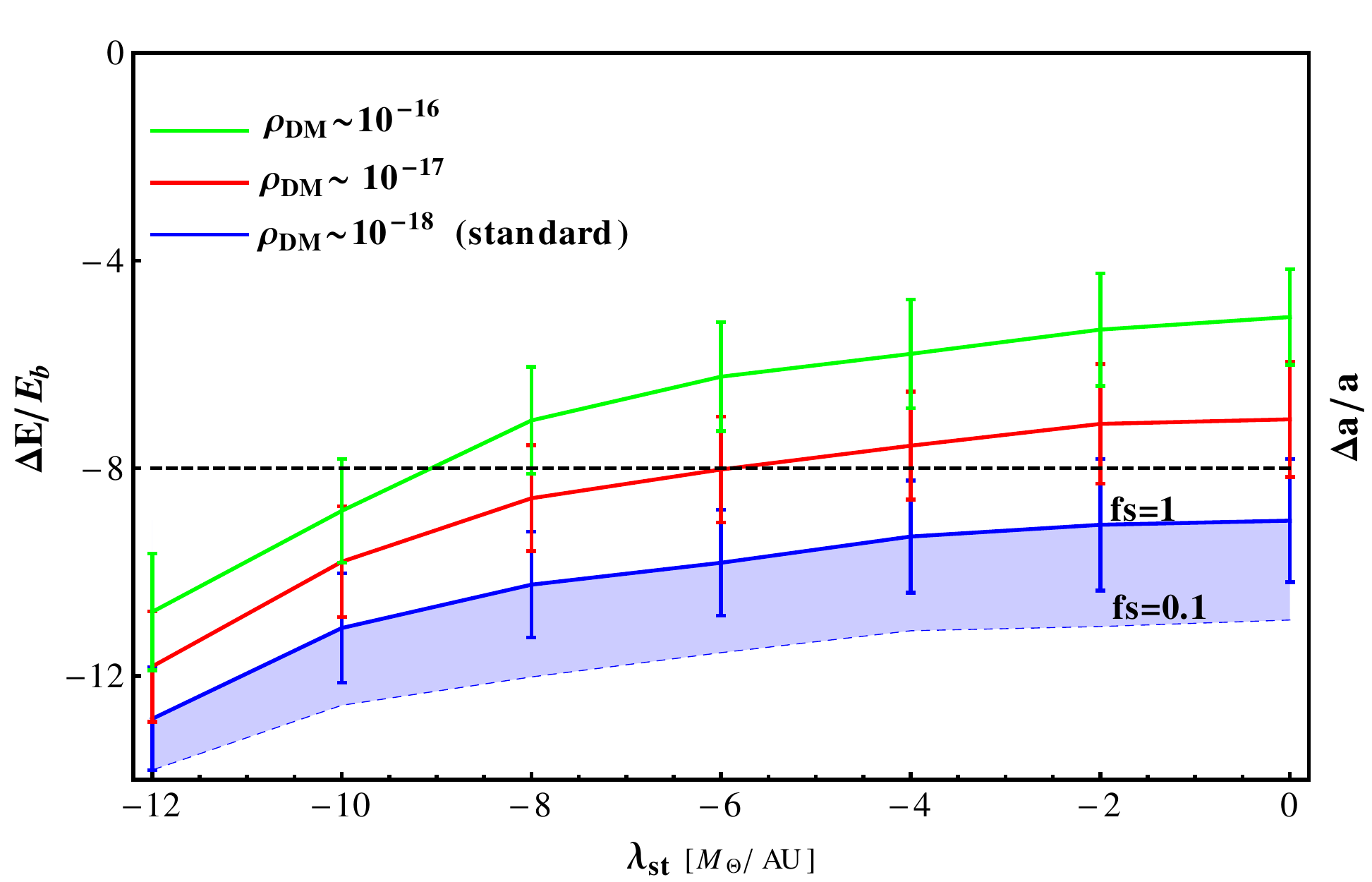}
                \caption{Neptune}
               \label{fig:neptune_stream}
        \end{subfigure}
       \caption{Constraints to linear density of dark streams due to encounters with the Earth-Moon System: We show the median
energy perturbation to (a) the Earth-Moon's orbit and (b) Neptune's Orbit, due to interactions with dark mini-halos as
a function of its mass and for three different values of the local $\textrm{DM}$ density, $\rho_{\rm DM}(\Msun/\AU^3)$. The bars correspond to the
first interquartile range for the Montecarlo realization, and the dashed line to the experimental uncertainty in the
semi-major axis of the orbit.}
\end{figure}

\begin{figure}[t!]
\centering
\includegraphics[width=0.5\textwidth]{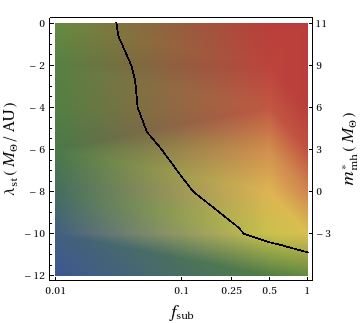}
\caption{Fractional injection energy, for the Earth-Moon system, as a function of $\rm f_{\rm s}$ and $\lambda_{\rm st}$. The solid line is where
the perturbation equals the observational limit, assuming the standard $\textrm{DM}$ local density. The right axis shows a possible relation with the
initial sub-halo mass, see section \ref{sec:initialminihalomass}.}
\label{fig:eamoon_str2}
\end{figure}

Now, lets turn our attention to the interaction of these systems with dark streams; the linear mass density values used here are in the range
$10^{-12}<\lambda_{\rm st}(\Msun/ \AU)< 10^0$. Figures~(\ref{fig:eamoon_stream}), and~ (\ref{fig:neptune_stream}), resumes the results. Contrary to
the point mass case, both systems could be significantly affected by the presence of dark streams. At the standard value of the local $\textrm{DM}$
density (blue line in the figures), the Earth-Moon system excludes the possibility of having the $\textrm{DM}$ distributed on streams with
$\lambda_{\rm st} \gtrsim 10^{-10} \Msun /\AU$, while the Sun-Neptune does not impose any restriction. On the other hand, if we consider higher values
for the local $\textrm{DM}$ density (red and green lines in figures) the constraints becomes more stringent. Both systems will be compatible with the
presence  of streams with linear density $\lambda_{\rm st} \lesssim 10^{-11} \Msun /\AU$, in the denser case. We initially assumed all the
$\textrm{DM}$ density is in
streams, $\rm f_{\rm s}=1$, once we relax this condition, (shaded blue area in the plots), there are several combinations of sub-structure fraction
and stream linear densities that are allowed by the observational constraints. The allowed and not allowed combinations are shown
in Figure(\ref{fig:eamoon_str2}): a density plot for the difference between the calculated fractional change to the semi-major axis of the orbit and
the observational restriction to the same quantity, the solid line indicates where the two quantities are equal, separating the region of the
parameter space that is excluded (yellow-red) from that allowed (green-blue).

A more realistic situation would be one in which a fraction of the mass is in the form of streams while the rest is in mini-halos. Since
the encounters with one or the other is independent. Then, in general the results will be set by the dominant specie. We can see from the plots that
the dominant contribution would come from the streams in most of the cases.

\paragraph{Evolution of the Astronomical Unit} There is evidence of some unexplained anomalies that concerns the astrometric data, including: the
Pioneer anomaly, the variations in the astronomical unit, the Earth flyby anomaly, and the increase in the eccentricity of the Moon's
orbit (all of them discussed in~\citep{2010IAUS..261..189A}) . The first one has been already explained by reanalyzing the thermal radiation pressure
forces inherent to the spacecraft~\citep{Rievers:2011mq}, thus it has been discarded the possibility that the presence of some $\textrm{DM}$is the
responsible for this anomaly. The second one remains unexplained, (see however~\citep{Iorio:2011zv}). The increase in the astronomical unit is
reported to be about $15\,cm\,yr^{-1}$~\citep{2010IAUS..261..189A}. We have applied our formalism of cumulative encounters with dark
sub-structure to the Sun-Earth system, to check if this anomaly could be explained by the dynamic perturbations due to the presence of sub-structure.
Assuming the standard local $\textrm{DM}$ density value, and regardless the sub-structure fraction, neither the perturbations due to mini-halos nor
due
to streams, for aboutthe lifetime of the Earth, would be energetic enough to explain this anomaly. If larger values of the local density were allowed,
the anomaly could be explained, however it would cause the secular evolution of other systems, like the Earth-Moon.

\subsection{Wide binaries in dSphs Galaxies}
\label{sec:Binc}
It is known that binary stars systems with separations as large as $\sim 0.1\, pc$ exist in the stellar halo of the
MW~\citep{1981ApJ...246..122B,1984ApJ...281L..41L}, but corroborating their existence in dwarf spheroidal galaxies is an
unachieved goal. The survival of these wide binaries against the interaction with the $\textrm{DM}$ has been discussed before. Those in the MW had
been used to
set strong limits to the MACHO's, (masive compact halos), mass\citep{Yoo:2003fr}. And now, the existence of those in the dSph's  had been proposed as
a discriminant for the $\textrm{DM}$ hypotesis~\citep{Hernandez:2008iq,Penarrubia:2010pa}. In particular in~\citep{Penarrubia:2010pa} it is argued
that
such kind of wide binaries should be disrupted due to interaction with $\textrm{DM}$ sub-halos. But, only catastrophic encounters, within the tidal
interaction approach, were considered. They also assume a sub-halo mass function like the one founded in $N$-body simulations. It is important to
recall that the formation process of wide binaries, and the determination of their lifetime is unclear~\citep{Moeckel:2010gt,Kouwenhoven:2010rc}, as
it is the assembling history of the dSph's. This uncertainties could play an important role, when estimating the dynamical perturbations that the
binaries suffer, due to the presence of dark sub-structure. According to previous studies, the existence of wide open binaries in dSphs would not be
compatible with the $\textrm{DM}$ hypothesis.

We study the disruption of wide binaries by means of their tidal interactions with two types of $\textrm{DM}$ sub-structure: mini-halos and streams. 
Considering only the effect of sub-structure that is bound to the dwarf halo, and not that bounded to the MW.  We can safely consider only the density
of $\textrm{DM}$ within the dwarf. The density of the Milky Way halo does not match that of the dSph's at the position of the binaries, otherwise the
dwarf and the binaries would be already unbounded~\citep{Read:2005zv}. We will follow the cumulative effect of interactions with sub-structures by
means of our Monte-Carlo experiments, this will automatically take into account the diffusive and the catastrophic regime. 

Such binaries could be present at a radius  $r\sim 0.1\, kpc$ from the center of the dSph galaxy, where a typical value for the
$\textrm{DM}$ density is $\rho_{\rm dm}\sim 2.2*10^{-17}\,\Msun/ \AU^3$, (assuming the galaxy is immersed in a NFW density profile with virial mass
$M_{\rm vir}=10^{9} \Msun$ and concentration $c=23.1$, as in~\citep{Penarrubia:2010pa}).\footnote{This particular radius is smaller, in
general, that the half-light radius of dwarf galaxies~\citep{Walker:2007ju}, and that the one assumed in~\citep{Penarrubia:2010pa}. We choose it this
way because at smaller radius, there are more stars, but also, the $\textrm{DM}$ density is higher, so the interactions could be more frequent.}
Dwarf galaxies have typical dispersion velocities of $\sigma \approx 10 {\rm km/s}$~\citep{Walker:2007ju}. We then assume the relative
velocity of encounters to follow a Maxwell-Boltzmann velocity distribution, as in the previous section, but with a dispersion of $\sqrt{2}\sigma \sim
14 \, {\rm {\rm km/s}}$.  Below, we present the results obtained by following the evolution of the binaries from their interaction with
dark sub-structure:
\begin{figure}[t]
  \begin{subfigure}{0.5\textwidth}
  \centering
 \includegraphics[width=\textwidth]{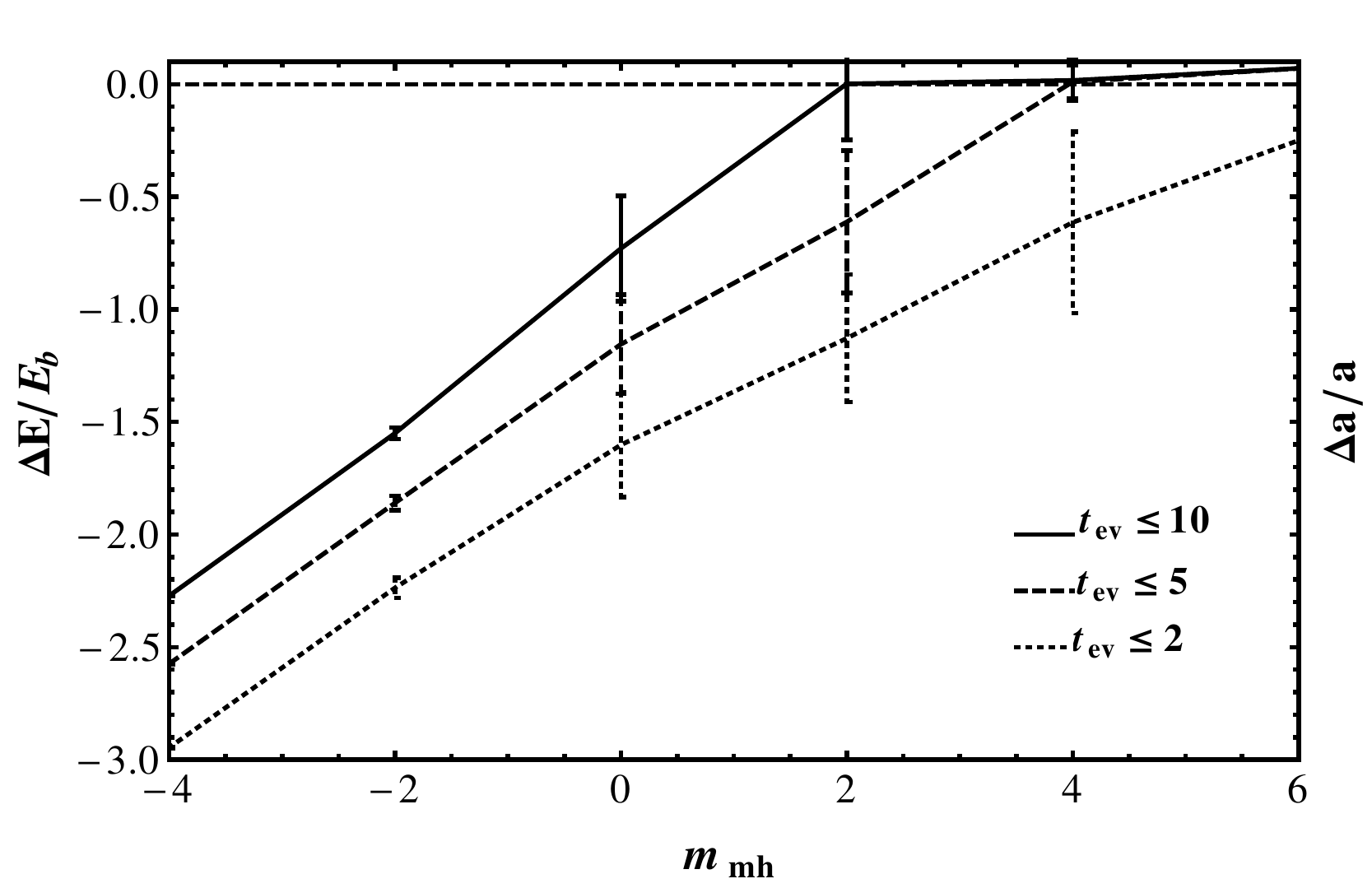}
 \caption{}
 \label{fig:binaria1}
 \end{subfigure}
\quad
\begin{subfigure}{0.5\textwidth}
 \centering
\includegraphics[width=\textwidth]{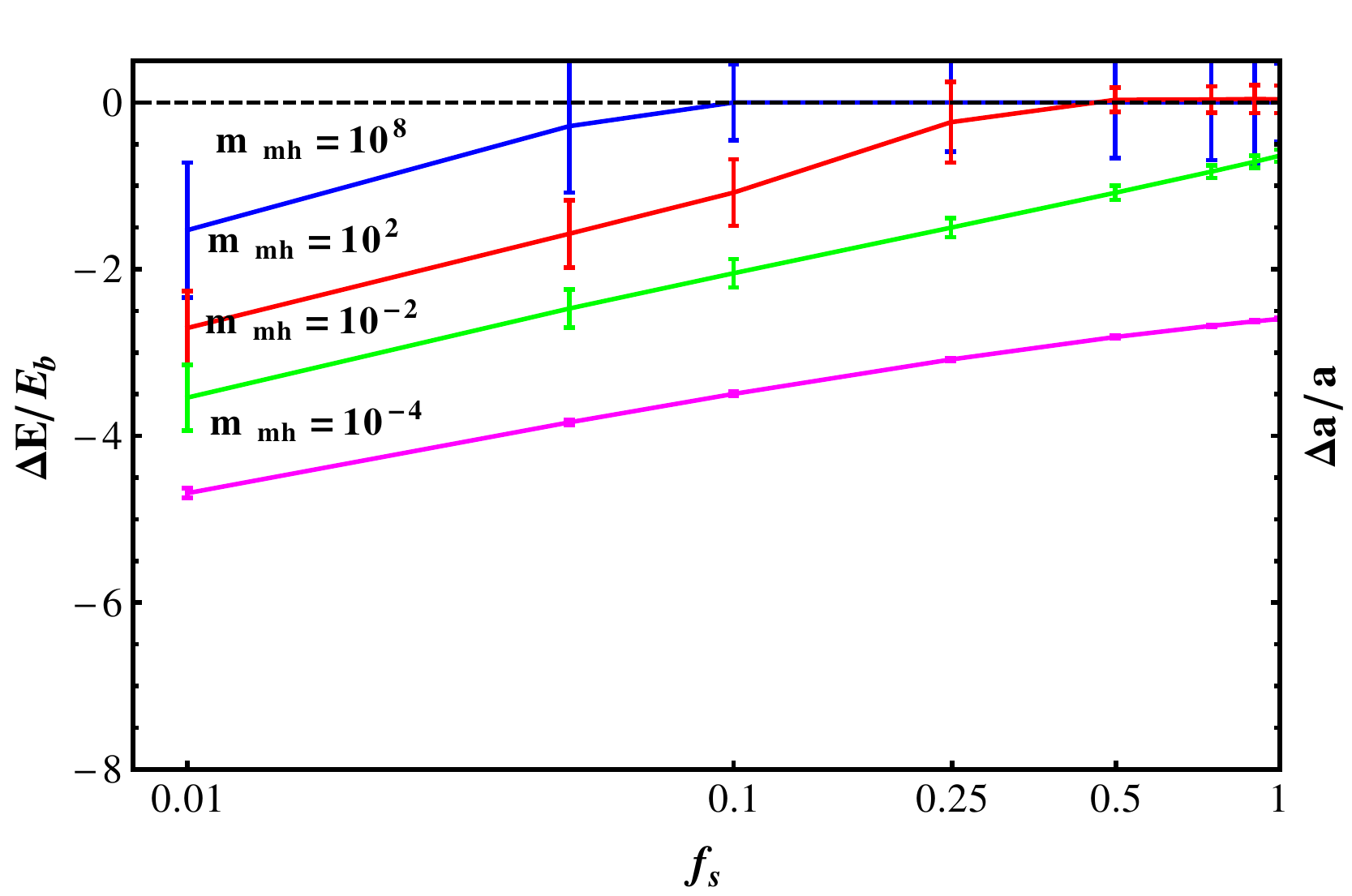}
\caption{}
\label{fig:binaria2}
\end{subfigure}
\caption{The median energy perturbation to a binary star system due to interaction with mini-halos. (a) As a function of the mini-halo mass,
$m_{\rm mh}(\Msun)$ , and for three different evolution times, $t_{\rm ev}(\rm Gyr)$. (b) As a function of the fraction, $\rm f_{\rm s}$, of the
mass density in mini-halos of different mass.}
\end{figure}

\begin{description}
 \item[Interaction with mini-halos] In Figure~(\ref{fig:binaria1}) we report the internal energy change, of the binary system, as a function of the
perturber mass in the range $10^{-4}<m_{\rm mh}<10^{8} \Msun$. For three different values of the evolution time: the solid line corresponds to the
typical value assumed in previous works ($t_{\rm ev}\sim 10\,\rm{Gyr}$), and the dashed, and dotted lines correspond to shorter times ($5 {\rm
Gyr}$ and
$2 {\rm Gyr}$). The line at $\delta E/E_{b}=0$ represents the limit at which the injection energy due to encounters equals the binding energy of the
binary, i.e. the moment at which the binary gets unbound. We can see that imposing a constrain to the lifetime of the binaries would impose a limit to
the maximum mass
of the sub-structure that could be present in the dSph's. However, this mass limit will also depend on the fraction of the mass that is in form of
mini-halos. In Figure \ref{fig:binaria2} we show the effect of varying the sub-structure fraction for selected values of the mini-halo mass. Note that
in the case of having massive perturbers, $>10^2 \Msun$ (the mass limit between the diffusive and catastrophic regime for tidal encounters set in
\citep{Penarrubia:2010pa}) it would be possible to unbound the binary, only if all the $\textrm{DM}$ were distributed in mini-halos of this mass, and
the evolution time is as large as $10\,Gyr$.The probability of having a massive perturber at short distance to the binary decrease as the impact
parameter has to be larger than for the less massive perturbers. Then it would be more probable to cause the disruption of the binary by the
presence of small mini-halos, than having a catastrophic encounter.

\item[Interaction with streams] For the interaction with streams we do the same as above, but, now as a function of the linear mass
density $\lambda_{\rm{st}}$. The results are shown in Figures~(\ref{fig:binaria3}) and ~(\ref{fig:binaria4}). Again, the binaries could be dissolved
for several combinations of $\lambda_{\rm st}$, $t_{\rm ev}$ and $f_{\rm s}$. Limits to the lifetime of the binaries could impose some limits to the
fraction of
the mass that can be composed of dark streams of a particular linear density. For instance, if we consider that binaries have been for about $2 Gyr$
under disruption, then there can not be present streams with linear densities greater that $10^{-6} \Msun/\AU$, composing more that the $10\%$ of the
total $\textrm{DM}$ density, otherwise the binaries would be disrupted. Similarly, if the binaries have been in the perturbation field for about $10
Gyr$, then there could be present streams with $\lambda_{\rm st}<10^{-10} \Msun/\AU$, or even $\lambda_{\rm st}<10^{-8} \Msun/\AU$, depending on value
of the $f_{\rm s}$ parameter. It is interesting to note that the limit to the linear mass density of streams in the later case is quite similar to
that imposed by the Earth-Moon system in section \ref{sec:SSc}.
\end{description}

A more general situation is that in which a fraction of the $\textrm{DM}$ density mass is distributed along streams, and other in mini-halos. In this
case the total energy perturbation will depend on the details of the parameters. One example is shown in Figure~(\ref{fig:binaria5}), where we compare
the perturbation produced by mini-halos with that produced by the streams. The comparison is made for two values of $m_{\rm mh}$ (red) and two values
of $\lambda_{\rm{st}}$ (blue). The intersection between blue and red lines shows the combination where one perturber type becomes dominant over to
the other. In addition, in the case of having a smooth halo component, the fraction of mini-halos and the fraction of streams are no
longer complementary, but the total perturbation effect can still be read from this plot.

\subsubsection{Comparison to previous work}

\begin{figure}[t]
  \begin{subfigure}{0.5\textwidth}
 \centering
 \includegraphics[width=\textwidth]{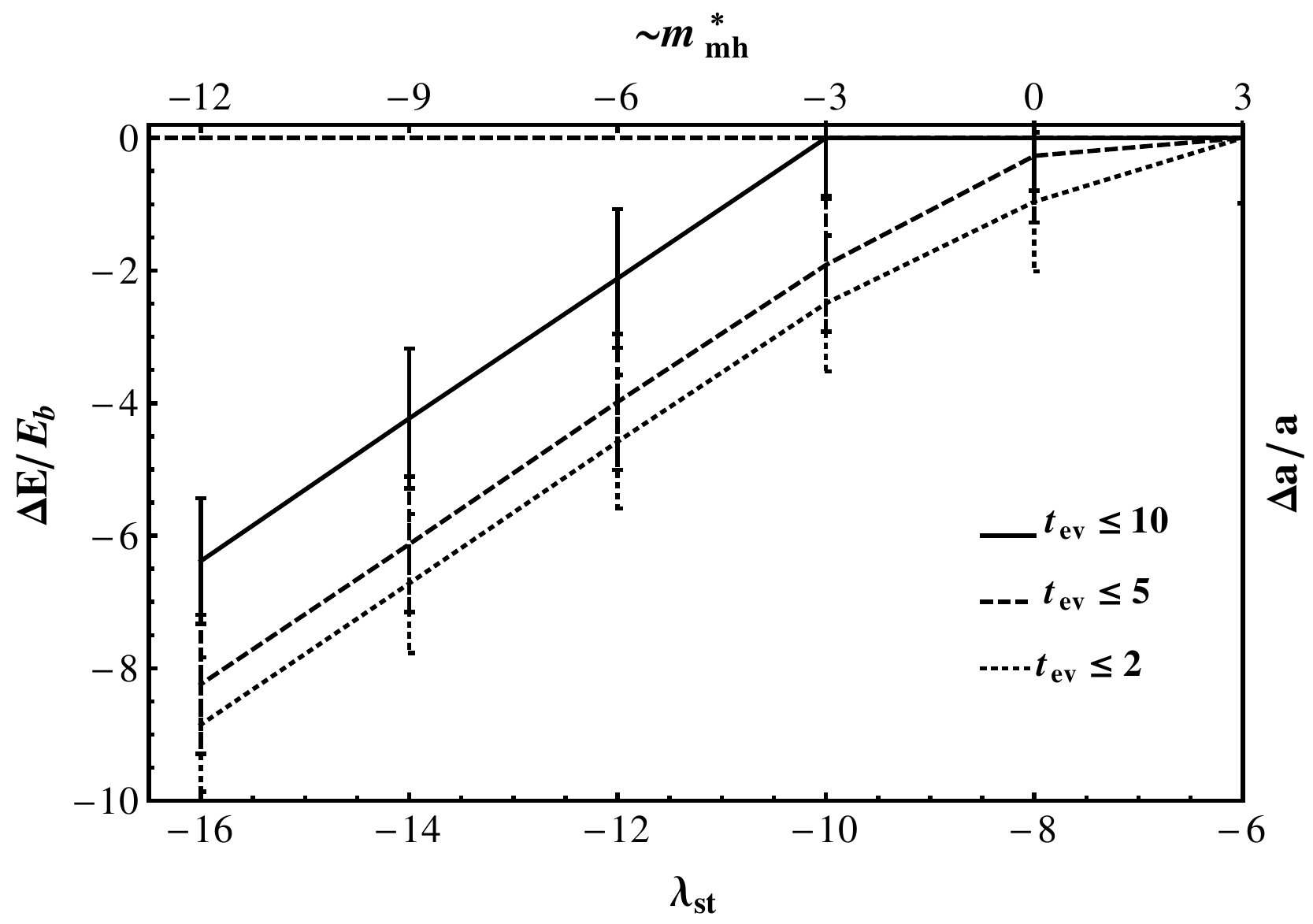}
 \caption{}
 \label{fig:binaria3}
 \end{subfigure}
\quad
\begin{subfigure}{0.5\textwidth}
 \centering
\includegraphics[width=\textwidth]{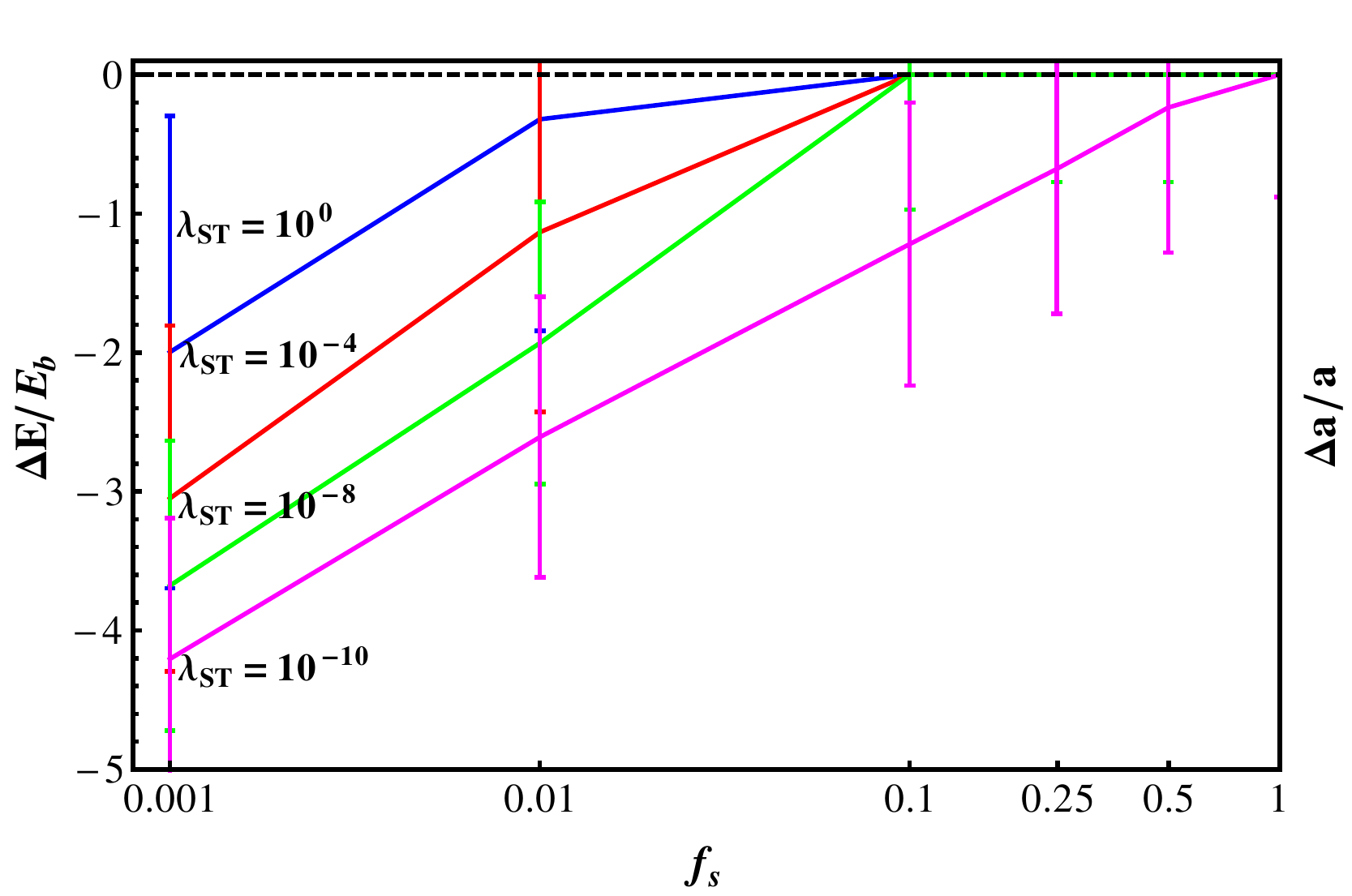}
\caption{}
\label{fig:binaria4}
\end{subfigure}
\caption{The median energy perturbation to a binary star system due to interaction with dark streams. (a) As a function of the stream linear mass
density $\lambda_{\rm st}$ and for three different evolution times , $t_{\rm ev}(\rm Gyr)$. (b) As a function of the fraction of the mass density in
streams, for four different linear mass density values. }
\end{figure}

\begin{figure}[t]
   \centering
 \includegraphics[width=0.5\textwidth]{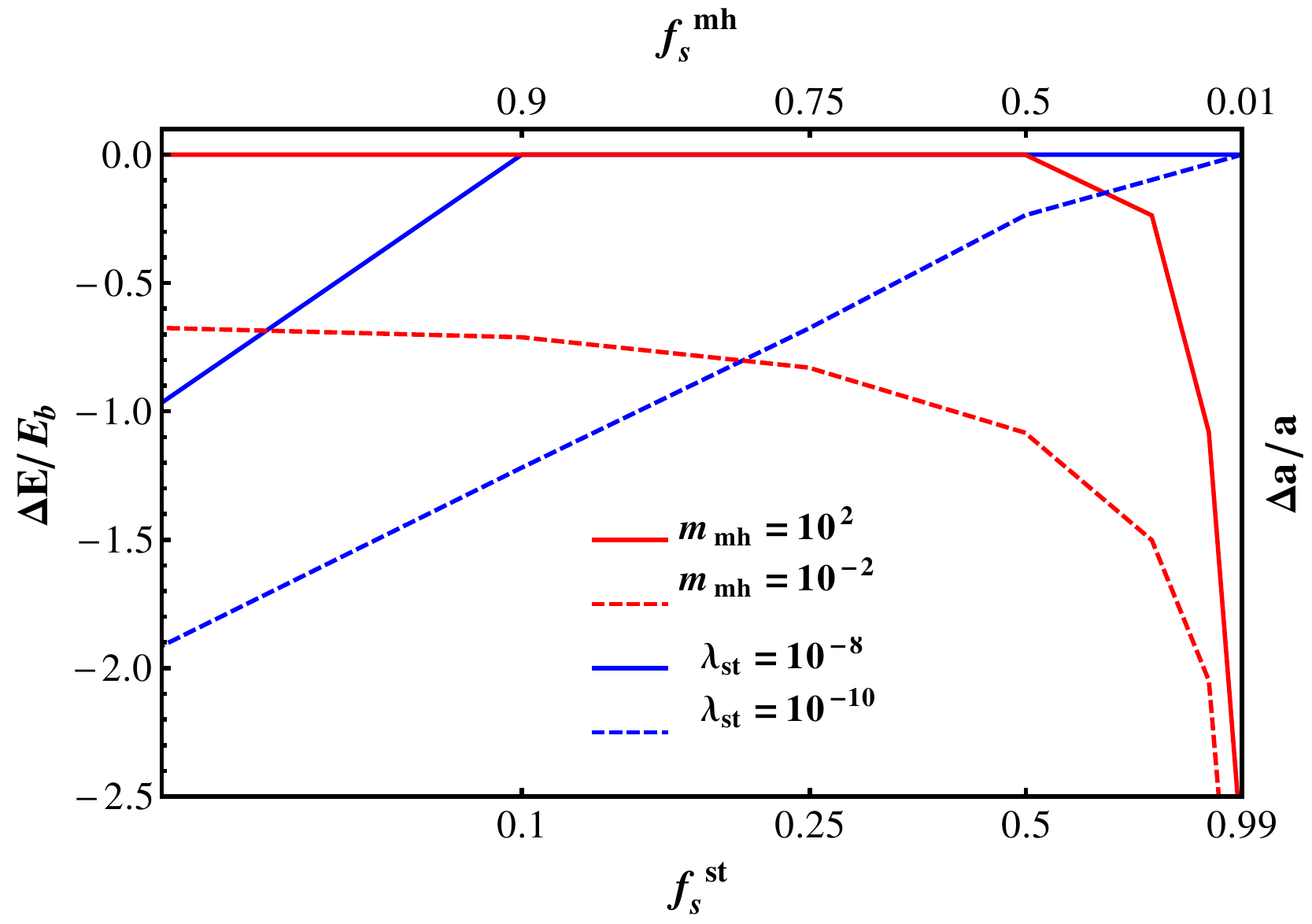}
 \caption{Binary perturbations due to the presence of mini-halos and streams. For selected values of the stream linear mass density
($\lambda_{\rm st}$, blue), an mini-halo mass ($m_{\rm mh}$, red), as a function of their corresponding sub-structure fraction. The intersection
between blue and
red lines shows the fraction where one perturber type becomes dominant over to the other.}
 \label{fig:binaria5}
\end{figure}

The disruption of Wide Binaries by their interaction with substructure was previously studied in~\citep{Penarrubia:2010pa}. HHere we resume the main
differences, and convergence points between both approaches. In this work we do not neglect a priori the contribution of the diffusive encounters to
the evolution of the binaries, but we follow the cumulative effect of encounters. In~\citep{Penarrubia:2010pa}  all the encounters considered are in
catastrophic regime. Tidal heating due to
diffusive encounters produces itself a secular evolution of the binary system, and it may modify the original binary configuration before a
catastrophic encounter could take place. The choice of encounter parameters in both approaches is also different. In~\citep{Penarrubia:2010pa}
these are tied to the results of N-body simulations, instead we sample the encounter parameters from random distribution functions that depends
only on the local $\textrm{DM}$ density. Note that the black solid line in figure ~\ref{fig:binaria1} recovers what should be expected from
~\citep{Penarrubia:2010pa}.  Our approach allow us to study the effect of the evolution time that binaries had been perturbed, $t_{ev}$. As the
evolution time decreases, the mass region for catastrophic encounters becomes smaller. Finally, we are able to study the dynamical effect due to the
presence of stream-like structures, figures~\ref{fig:binaria3} and \ref{fig:binaria4}; and mixed populations of mini-halos and
streams, figure~\ref{fig:binaria5}.  So far the only missing point here is to implement a relation between the perturber mass,
$m_{\rm mh}$ , and the substructure fraction, $f_s$ , which in principle can be taken from the extrapolation of N-body simulations at larger
scales~\citep{Springel:2008cc}.  Although this extrapolation may be verified in future studies, we want our approach remain as independent as
possible. However, we expect that
the result of including a substructure mass function in our analysis will fit in the limit cases we have already considered. Overall, we find that
there is a large region in the encounter parameter space that produces secular evolution of the binary systems, then the final distribution of binary
separations could not be independent of the mass and abundance of substructure, making this test not to be a sharp discriminant of the DM
hypothesis as claimed in~~\citep{Penarrubia:2010pa}. However it would be very useful to constrain the properties of the substructure, specially when
combined with other astrophysical constraints.

\section{A possible relation between $\lambda_{\rm st}$ and $m_{\rm mh}$}
\label{sec:initialminihalomass}
The actual constraints for the streams like structures could be interpreted in terms of their original mini-halo mass if we had information
about the disrupted time ($t$) and how the disruption was. These two aspects are not easy to determine. We propose a very simple approach to establish
this connection. Estimate the mass per unit length of a resultant stream, $\lambda_{\rm st}$, by
computing
first its length $L$ with the relation $L \approx \sigma_{\rm mh}\,t$ \citep{Schneider:2010jr}, which assumes that the particles moves with the
velocity dispersion of the progenitor mini-halo. For simplicity, we will consider the mini-halos had an initial NFW density profile, and so the
maximum velocity
dispersion
can be inferred from the virial mass~\citep{Klypin:1997fb} through
\begin{eqnarray}
 \sigma_{0}^2 &\approx& 0.472 V_{\rm max}^2 \\ \nonumber
 V_{\rm max}^2 &=& \dfrac{G\,M_{\rm vir}}{2 r_{s}} \dfrac{0.432}{Log[1+c]-\dfrac{c}{1+c}}
\end{eqnarray}

where $r_s$, and $r_{\rm vir}$ are the scale and virial radius, respectively. Use the following definitions: $r_{\rm vir}=(M_{\rm vir}/(4 \pi/3
\rho_{\rm cr} \Omega_{\rm m} \delta_{\rm th}))^{1/3}$ with $\rho_{\rm cr}=270.5 \Msun/Kpc^3$, the critical density of universe and $\delta_{\rm
th}=340$ the  predicted over-density of a collapsed object according to the top-hat collapse model for the $\Lambda-\textrm{C$\textrm{DM}$}$ model
with $\Omega_{\rm m}=0.3$); and, the concentration for mini-halos  $c=r_{\rm vir}/r_s=12*(M_{\rm vir}/h^{-1}\,10^{12})^{-0.12}$~\citep{Klypin:2010qw}.
Given the above relations, the resultant velocity dispersion for the halo mass in the range $10^{-6}<m_{\rm mh}(\Msun)<10^0$ results to be
$10^{-4}< \sigma_{\rm mh}\,({\rm km/s}) < 10^{-2}$, so after an evolution of a Hubble time, the length of these structures would be $7
<L\,(pc)<370$. Then, using a simple linear fit the relation between the mini-halo mass, and the linear mass density could be roughly set to
\begin{equation}
\lambda_{\rm st}\approx 10^{-8} m_{\rm mh}^{0.7}.
\label{eq:lambdarelation}
\end{equation}
For instance, an initially $10^{-6} \Msun$ mini-halo mass would be associated with a stream of
linear density $10^{-13}\Msun/ \AU$, which according to Figure~(\ref{fig:eamoon_stream}) would be perfectly allowed by the dynamics in the
Solar System. For comparison we have scaled our results in Figures~(\ref{fig:eamoon_str2}), and~(\ref{fig:binaria3}), using the relation
in~\eqref{eq:lambdarelation}. However, in order to have a more accurate connection between $\lambda_{\rm st}$ and $m_{\rm mh}$ an statistical study
for the interaction between mini-halos and perturbers should be performed, but this is out of the scope of this work.

\section{Discusion and conclusions}
\label{sec:conclusions}
We have estimated the dynamical effect that sub-structure in the galactic halo may trigger on the evolution of the planet 
orbits in the Solar System; the Sun-Neptune and Earth-Moon orbits specifically. The effect has been compared to the current precision in the
determination of their semi-major axis (known with millimeter accuracy for the Earth-Moon thanks to the APOLLO-LLR intiative
\citep{2012CQGra..29r4005M}), finding that a local
$\textrm{DM}$ distribution completely populated by stream like structures with linear density of $ \lambda_{\rm st} \gtrsim 10^{-10} \Msun/
\AU$ in the solar neighborhood, can cause a secular evolution to the Earth-Moon system that would be above current measured limits
(though Neptune orbit allows a higher value). We have also shown the allowed, and not allowed, parameter regions, in a  ``linear
density-sub-structure fraction'' space, for dark streams based on the current precision of the Earth-Moon distance measurements.

Here it is worth to emphasize the importance of the local $\textrm{DM}$ density value in our study. The typical approach to constrain its value is
based on dynamical studies, however, many astrophysical uncertainties are associated. According to different authors the actual limits varies from
$\approx 10^{-18}
\Msun/\AU^3$~($(0.2-0.85)\,\textrm{GeV}\,
\textrm{cm}^{-3}$)~\citep{Garbari:2012ff,Widrow:2005bt,Catena:2009mf,Iocco:2011jz,Salucci:2010qr,deBoer:2010eh}. Most of these studies are based on
the rotation curve of the galaxy, and are subjected to the systematic errors associated to the unknown shape of
the galactic halo. Other determinations of the density are local and in principle the systematic errors can be eliminated~\citep{Garbari:2012ff}.
The actual data quality still needs to improve in order to get better constraints on the local $\textrm{DM}$ density. The GAIA astrometric mission
will
improve these determinations~\citep{Papastergis:2011xe,Brown:2005xz}. 

It is important, as well, to say that all these determinations only give information about the mean value, at the scale of 100's to 1000's of parsecs,
but says nothing about the way the $\textrm{DM}$ is distributed at smaller scales. Indeed, the density at the Solar System scales could show important
over(under)-densities regardless of the mean value at larger scales. In this work we used different values of the local $\textrm{DM}$ density, that in
some cases are out of the range allowed by the current observational uncertainties. This was done in order to independently explore the space
parameter. At the end, our study seems to indicate that if the Solar System
$\textrm{DM}$ were mainly distributed in mini-halos, the density could be an order of magnitude greater than the standard value, without being in
contradiction with the Solar System dynamics and the current observational constraints described just above, but not if it is mainly distributed along
streams.

We have also applied our formalism to determine if the presence of Dark Substructure in dSphs Galaxies could be in conflict
with the presence of wide open binaries inside dSphs. The case of cumulative encounters, the diffusive regime, and the interaction with streams 
has not been considered elsewhere. We have demonstrated that catastrophic encounters are not, necessarily, the driven mechanism of destruction of wide
binaries. The effect of cumulative encounters with small perturbers can cause secular evolution as well. Our results indicate that to have better
constraints on the lifetime of the binaries could help to impose constraints to the properties of the dark sub-structure. But, as there are many
parameter combinations, ($m_{\rm mh}$, $\lambda_{\rm st}$, $f_{\rm s}$, $t_{\rm ev}$), for which it is possible to cause secular evolution of the
binary system, or even dissolve it; we consider that the existence of this kind of binaries is not a sharp discriminant of the $\textrm{DM}$
hypothesis.
However, to get a good characterization of their distribution would be helpful to constrain the properties of the dark sub-structure, specially when
combined with other astrophysical restrictions. 

Also, by relating the linear density of streams with a mini-halo progenitor mass (a simple approach) we conclude that the typical cut-off in the mas
power spectrum, i.e. the mass of the smallest structures ever formed in the Universe $m_{\rm min} \sim 10^{-4}-10^{-12} \Msun$
\citep{Green:2003un,Profumo:2006bv}, will correspond to streams with linear densities of $ \lambda_{\rm st} \sim 10^{-11} - 10^{-17} \Msun/\AU$. Such
values are compatible with the dynamics of the solar system (a similar conclusion was obtained when we consider the mini-halos as a point mass
perturbers), and would be compatible also with the existence of wide binaries in dSph's, if they were to be detected. On the other hand, for streams
with $\lambda \approx 10^{-10}$--$10^{-8} \Msun/\AU$ ($m_{\rm min} \sim 10^{-3}-10^{0} \Msun$), the fractional energy imprinted to the binary stars
would cause their disruption, and the corresponding effect to Earth-Moon system would be just below (or above) the observational restriction. That is,
if wide binaries in dSph's with separations of $a\sim 0.1 pc$ were not observed, it could be due to the presence of streams with  $\lambda_{\rm st} >
10^{-10} \Msun/\AU$ ($m_{\rm min} \sim 10^{-3}$), but the presence of such kind of structures in the solar neighborhood would (likely) be incompatible
with the Earth-Moon dynamics (although it depends in the sub-structure fraction). This is an example of how the combined restrictions, from different
astrophysical systems, could be revealing the sub-structure mass function cut-off scale.

Finally, we want to make some comments regarding the direct, and indirect $\textrm{DM}$ detection methods. First, it is well known that  the presence
of dark mini-halos, or streams, in the solar neighborhood, could affect the event rate and energy spectrum, and consequently the exclusion limits set
by current experiments. From our study, we can say that the dynamics of the Solar System is not significantly affected by the presence of both kind of
sub-structures, mini-halos and streams (not even in the idealized case when the sub-structure fraction is equal to one). This rise up the possibility
of being in a over-dense region and suggest that $\textrm{DM}$ halo models that includes the presence of $\textrm{DM}$ sub-structure should be used
when interpreting experimental results. However, there is also the possibility to be in a under-dense region, not distinguishable from the standard
case by dynamical studies, but with important consequences for the direct detection methods.

On the other hand, the presence of dark sub-structure in galactic halos, could also affect the interpretation of the results for
$\textrm{DM}$ indirect detection experiments, because in some cases it is subjected to the use of a boost factor mechanism, usually
associated to the Sommerfeld enhancement, that is velocity dependent.\footnote{If Dark matter is composed by WIMPs particles, the
detection of their products after annihilation offers an opportunity to indirectly infer the existence of this kind of particles,
the Fermi and PAMELA experiments are two great experiments that has opened the possibility of explaining their electron and
positron excess  through the $\textrm{DM}$ hypothesis~\citep{Adriani:2008zr,FermiLAT:2011ab}. Though the signal could also
be attributed to other, yet unresolved, astrophysical sources.} As mini-halos and streams are, in general, kinematically colder
than a soft galactic halo, then, the limits to the sub-structure properties set in this work could be useful also to interpret the
results of $\textrm{DM}$ indirect detection experiments .

\acknowledgments 

 A.X.G-M. and O.V. thank DGAPA-UNAM grant No. IN115311  and CONACyT CB-128556. A. X. G-M. work was supported by a CONACyT PhD fellowship.

\bibliographystyle{JHEP}
\bibliography{AXGonzalez-Streams}

\appendix

\section{Corrections for Spherical models.}
\subsection{Correction for extended perturbers.}
\label{sec:appendix1}
In this section we present the behavior of the $U$ function, in equation \eqref{eq:deltaEpoint}, that accounts for the extension and structure of the
dark
mini-halos. For this we have assumed the mini-halos have a NFW mass profile, normalized to the virial mass, of the form:
\begin{equation}
 \mu (r/r_{s})=\dfrac{1}{\log(1+c)-\dfrac{c}{1+c}}\,\left(\log\left[1+r/r_s\right]-\dfrac{r/r_s}{1+r/r_s}\right),
\end{equation}
where $c=r_{\rm vir}/r_s$ is the concentration; $r_s$ and $r_{\rm vir}$ are the scale and virial radius respectively.
Following~\citep{1985ApJ...295..374A}, and given that the mass profile is normalized to the virial mass, we will define the $U$ function as:
\begin{subnumcases}{U(\xi)=}
\int_{1}^{\infty}{\mu(p\,\xi)\, \xi^{-2}(\xi^2-1)^{-1/2}} d\xi \quad \mbox{for}\quad p/r_s< c \nonumber \\
\int_{1}^{\infty}{\xi^{-2}(\xi^2-1)^{-1/2}} d\xi = 1\quad \mbox{for}\quad p/r_s > c
 \end{subnumcases}

\begin{figure}
   \centering
   \includegraphics[width=0.5\textwidth]{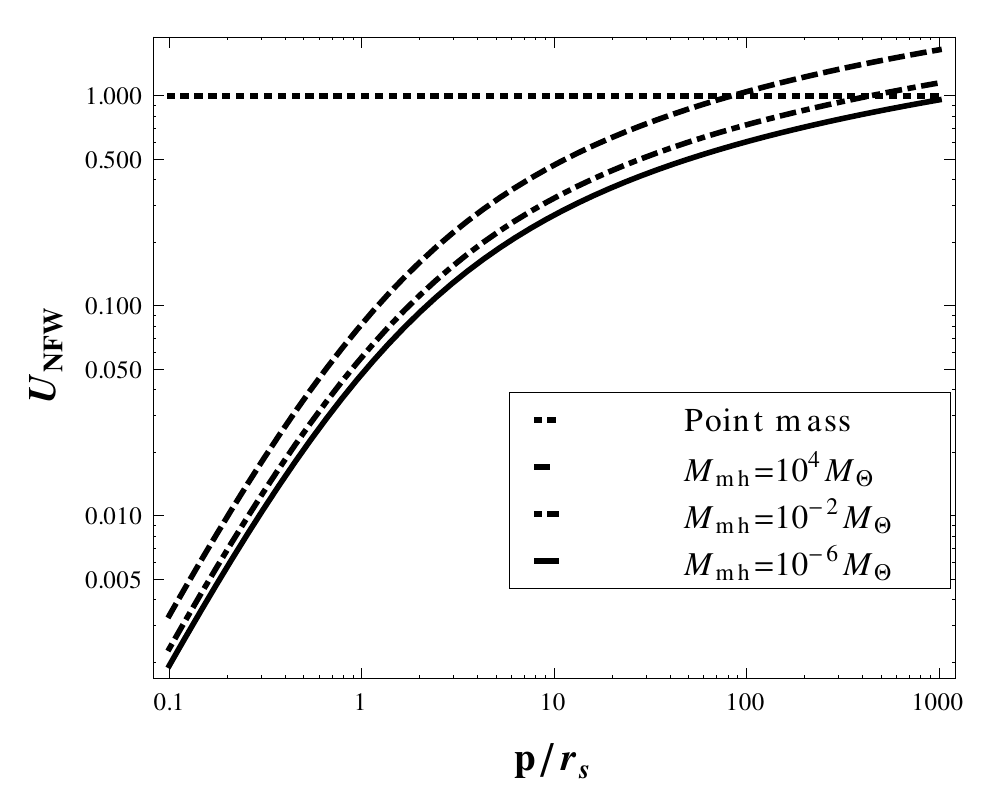}
   \caption{Structure function, $U$, for halos with a NFW density profile, as a function of their mass. For comparison we added a line for the point
mass case ($U_{\rm NFW}=1$), to see the scale at which this becomes valid.}
\label{fig:Ucorr}
\end{figure}

After integration, the $U$ function takes the form
\begin{subnumcases}{U(\xi)=}
 \dfrac{\dfrac{{\rm ArcSec}[p]}{\sqrt{-1+p^2}}+\log\left[\frac{p}{2}\right]}{\left(-\frac{c}{1+c}+\log[1+c]\right)} \quad
\mbox{for}\quad p/r_s< c \nonumber \\
1  \quad \mbox{for}\quad p/r_s > c,
 \end{subnumcases}
and it is ploted in Figure, \ref{fig:Ucorr}, for different values of the mini-halo mass. The $U$ function crosses the point mass approximation at
$r/p=c$ for the different masses. For comparison we added a line for the point mass case ($U_{\rm NFW}=1$), to see the scale at which this becomes
valid. We have seen this correction is less important for small mini-halos.

 \subsection{Corrections for non-distant tide encounters.}
\label{sec:appendix2}
In this section we deal with encounters where the distant tide approach is not applicable. In particular, this happens when
we study the interaction of wide binaries with point mass (or spherical) dark sub-structure present in the MW satellites.
We will use the this system to state the problem, and the solution, but this can be applied to other systems for which the distant
approximation does not applies. 

\begin{figure}[htb]
  \centering
  \begin{minipage}[c]{0.5\textwidth}
  \begin{subfigure}{\textwidth}
  \centering
      \begin{tabular}{|c|c|}
 \hline
  $m_{\rm mh}(\Msun)$&$p/a$\\
  \hline
  $10^{-6}$ & 0.33 \\
  $10^{-4}$ & 1.55\\
  $10^{-2}$ & 7\\
  $1$ & 33\\
  $10^2$ & 155\\
  $10^6$ & 335\\
  \hline
 \end{tabular}
 \caption{}
 \label{fig:bintab}
 \end{subfigure}
  \end{minipage}
  \quad
  \begin{subfigure}{0.45\textwidth}
   \centering
   \includegraphics[width=\textwidth]{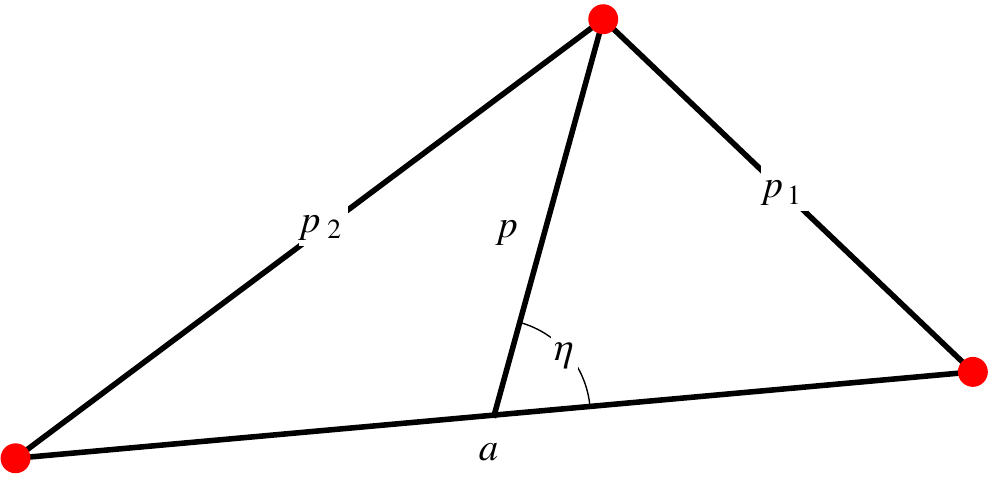}
    \caption{}
    \label{fig:bincorr}
 \end{subfigure}
\caption{(a) Shows the factor $p_{\rm mean}/a$ for different values of the mass perturber. Values close to one means that the
distant approximation, in the calculation of the dynamical effect of encounters between a target and a population of mimi-halos,
is not applicable. (b) Diagram that shows the definitions used to calculate the correction for non-distant encounters with
mini-halos. }
\end{figure}

The distant tide approximation is valid when the size of the target system is much smaller than the impact parameter. Te determine
whether this condition is satisfied, or not, we consider the characteristic of the wide binaries and $\textrm{DM}$ sub-structure we
used in Sec.\ref{sec:Binc} ($\rho_{\rm dm}=2.2 10^{-17} \Msun/\AU^{3}$, a=$0.1pc$, $M_{b}=1\Msun$, etc.). We can compare the binary
separation with the mean impact parameter for the encounters, $<p>=\Gamma(1/3) \, (36 \, \pi \, \rho_{\rm dm}/m_{\rm mh})^{(-1/3)}$
(defined from the nearest neighbor distribution), for different values of the perturber mass. Table \ref{fig:bintab} shows this
comparison: values lower or close to one implies that distant tide approximation is not applicable.
For the less massive perturbers, most of the encounters could not be treated in the distant tide approximation. Since the
sampling of impact parameters is random, we do not know, a priori,  which of the encounters could be treated in the distant
approximation and which not. In the practice we have to make this comparison for each encounter, then we decide if it is
appropriate to use the distant tide approximation, or to directly calculate the velocity kick to each component of the binary, and
calculate the input energy as we describe in the following paragraphs. 

For each component of the binary the kick in velocity due to an encounter, in the non-distant tide approach, is given by 
 \begin{equation}
  \Delta \mathbf{v}_{i}= \dfrac{2 G M_{p}}{v} U(p_{i})\dfrac{\hat{ \mathbf{p}_{i}}}{p_{i}}
 \end{equation} where $U(p)$ is a function that depends on the perturber density profile, and tends to one when the impact
parameter ($p$) is larger compared with the size of the perturber (this correction is described in detail in
\citep{1985ApJ...295..374A}, and in appendix~\ref{sec:appendix1}). The subscript $i$ refers to the $i$-th component of the binary. The differential
energy per
unit mass transferred to the binary system is 

 \begin{equation}
  \Delta E= 0.5 \Delta v_{\rm diff}^{2} +  \mathbf{v} \cdot \Delta  \mathbf{v}
  \label{eq:deltaE}
 \end{equation}
where $\Delta v_{\rm diff}= \left| \Delta \mathbf{v}_{2}-\Delta  \mathbf{v}_1\right| $, is the vector difference of the individual
contributions of both components of the binary system. 

In our approach the magnitude of the impact parameter is sampled from a nearest neighbor distribution, so we have to
assume respect to what point point of the binary it is referred, by simplicity we will assume it is respect to the center of
mass, so that the individual impact parameters to the binary components are given by:

\begin{eqnarray}
 p_{1}^2=p^2+(a/2)^2-p\,a\,\cos{\eta} \nonumber\\
 p_{2}^2=p^2+(a/2)^2+p\,a\,\cos{\eta}
\end{eqnarray}
where $\eta$ is the relative direction between the imaginary line that connects the binary, and the one that draws the impact
parameter direction (we will sample this angle from a uniform distribution between $0$ and $\pi$, since it is equally probably to
have any orientation of the binary). Given this geometry we rewrite eq. \eqref{eq:deltaE} in terms of the impact parameters
and the binary separation.

\begin{eqnarray}
\Delta E&=& 0.5 \left(\dfrac{2 G M_{p}}{v}\right)^2 \delta v_{\rm tot}^2 \;\;\;\;\; \mbox{where} \nonumber\\
\delta v_{\rm tot}^2 &=& \delta v_{1}^2 +\delta v_{2}^2 - 2 \delta  \mathbf{v}_1 \cdot \delta  \mathbf{v}_2 \nonumber \\
& =& \left(\dfrac{U(p_1)}{p_1}\right)^2+\left(\dfrac{U(p_2)}{p_2}\right)^2-2 \dfrac{U(p_1)\,U(p_2)(p_1^2+p_2^2-a^2)}{p_1^2\,p_2^2}
\label{eq:deltaEtot}
\end{eqnarray}

This last expression is used whenever the condition of distant tide encounters, in our analysis, is not satisfied.  Note we have
neglected the linear velocity term in eq \eqref{eq:deltaE}, beacuse we expect it approaches to zero when $\Delta E$ is
summed over all the encounters as a result of  symmetry (See~\citep{1985ApJ...295..374A}).

\section{Stream models}
\subsection{Derivation of tidal perturbation.}
\label{sec:appendix3}
In this appendix we derive analytical expressions for the force and tidal fields produced by the stream models we have
considered.  All of them present axial symmetry. Our simplest model is a 1-dimensional mass distribution with linear mass density
$\lambda_{\rm st}$ (S-1D). Then we consider a stream with finite cross section of radius $R_o$ and constant cross--sectional density
$\Sigma_o$ (S-CD). Finally, we consider streams whose cross-sectional densities vary as power laws with central cores
(S-Core), or cusps (S-Cusp):
\begin{equation}
\Sigma\left(R;R_o,\alpha,\gamma\right) = \Sigma_o\, \left(R/R_o\right)^{-\alpha}\,[1+\left(R/R_o\right)]^{\alpha-\gamma}, 
\end{equation}
where $\Sigma_o$ is a scaling density parameter, $R_o$ a characteristic radius and $\alpha$, $\gamma$ are dimensionless exponents
that satisfy:
\begin{equation}
0\le\alpha<2, \,\, \gamma>2.
\end{equation}
The restriction $\alpha\ge 0$ is to avoid an inverted density profile at the center, while $\alpha<2$ avoids an infinite mass
singularity there ($0<\alpha<2$ produces a finite mass singularity). The restriction $\gamma>2$ is to have a finite mass per
unit length. When $\alpha=0$ we have the flat core cases, whereas for $\alpha>0$ we have the cusp cases.

\subsection*{Force field}
The force field can be obtained using the integral Poisson equation:
\begin{equation}
\int\nabla\cdot{\bf F}\, dV = -4\,\pi\, G\,M.
\end{equation}
Taking advantage of the axial symmetry of the problem, the resulting forces are given by:
\begin{eqnarray}
F^{S-1D}\left(R\right)     =& -\dfrac{2G\lambda_{\rm st}}{R},
\label{eq:stream1D}\\
F^{S-CD}\left(R\right)    =& -F_o\, \dfrac{{\left(Min[\eta,1]\right)}^2}{\eta},
\label{eq:stream2D}\\
F^{S-Core}\left(R\right) =& -F_o\,
\dfrac{\eta^2-\gamma\eta\left(1+\eta\right)+{\left(1+\eta\right)}^\gamma-1}{\eta{\left(1+\eta\right)}^\gamma},
\label{eq:streamcore}\\
F^{S-Cusp}\left(R\right) =& -F_o\, \dfrac{1-{\left(1+\eta\right)}^{2-\gamma}}{\eta} \,\,\,\,\,\, \left(\alpha=1\right),
\label{eq:streamcusp}
\end{eqnarray}
where we have introduced the dimensionless length $\eta = R/R_o$ and $F_o\equiv 2G\lambda_{\rm st}/R_o$, with $\lambda_{\rm st}$ equal to the total
mass per unit length (integrated across the stream cross section). All forces are orthogonal to the stream and directed
towards it. For the cusp case there is no general analytical expression, so we show the special $\alpha=1$ case, which is the only
cusp model we will consider from now on. 

\begin{figure}
   \begin{subfigure}{0.49\textwidth}
\centering
\includegraphics[width=\linewidth]{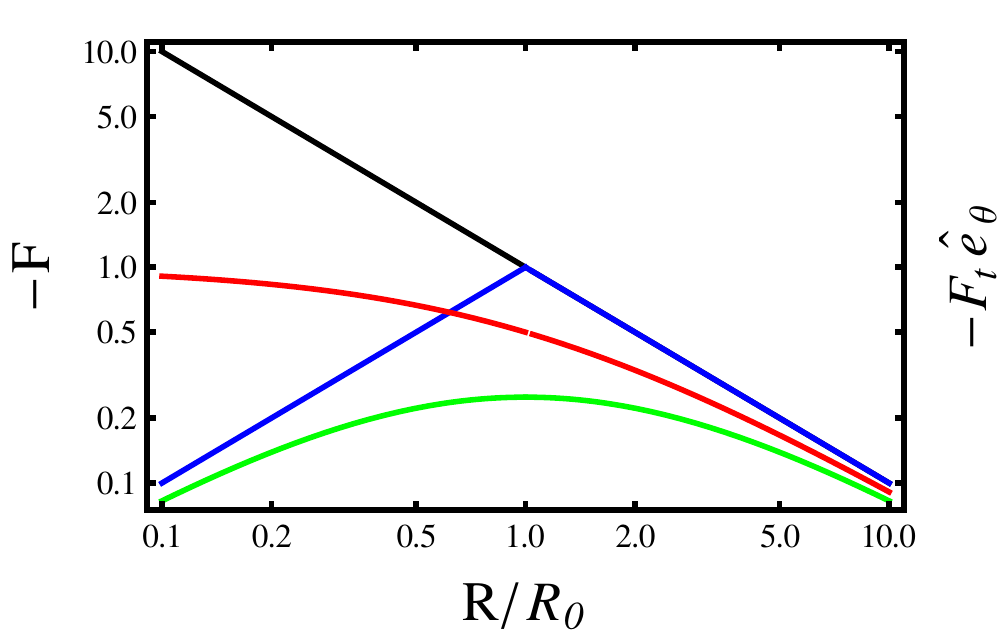}
\caption{Magnitude of the force produced by our stream models as a function of distance to the stream central axis, $R$. It coincides with the tidal
force in the $\theta$ direction, equation \eqref{eq:tidalforce}.}
\label{fig:forceFils}
\end{subfigure}
\quad
\begin{subfigure}{0.47\textwidth}
\centering
\includegraphics[width=\linewidth]{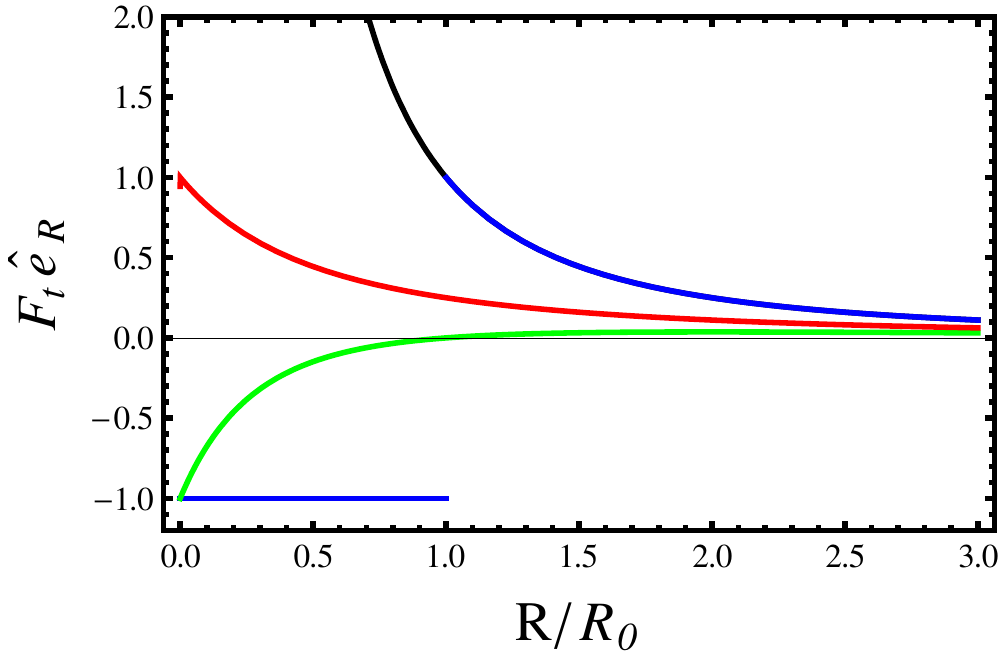}
\caption{Magnitude of the tidal force, in the $R$ direction, produced by our stream models as a function of distance to the stream central axis, $R$.}
\label{fig:tidalforceFils}
\end{subfigure}
\caption{The curves are for the 1-D (black), constant density (blue), core (green) and cusp (red) models. In each case, we set $\lambda_{\rm st}$,
$F_0$ and $\zeta_{0}$ to the unity. For the core and cusp models $\gamma=3$ is assumed. }
\label{fig:Fils}
\end{figure}

Figure~(\ref{fig:forceFils}) shows these forces, for streams with $\lambda_{\rm st}=1$, or $F_0=1$, accordingly; and $\gamma=3$ for the core and
cusp models is assumed. We see that the S-1D and S-CD models coincide for $R>R_o$, but whereas the former diverges toward the center, the latter
converges linearly to zero. The core model goes also to zero at the center, while the cusp model does not and so the force is discontinuous at the
stream axis. For large $R$ all models coincide, as they should, since they become indistinguishable.

\subsection*{Tidal force field}

The components of the tidal force produced by a force field ${\bf F}$ on a target of size $d {\bf r}$ at position ${\bf r}$, in
the linear approximation, (we use general coordinates $x^{\alpha}$ and Einstein's sumation convention), is given by :
\begin{equation}
{F_{t\,x_{\alpha}}} = -\dfrac{\partial { F}_\alpha}{\partial x_\beta}
dx^{\beta} =\tau_{\alpha\,\beta} dx^{\beta} 
\label{eq:tidaltensor}
\end{equation}
The tensor $\tau$, a symmetric and second rank tensor, is the Jacobian of the gravitational force field ${\bf F}$, and the
negative of the Hessian matrix, $H$, of the scalar potential function $\Phi$. \footnote{The vector
${\bf r}$ spans the range across where the differential effect of the tide is to be calculated.} The tensor $\tau$ can be written
in any curvilinear
coordinates by means of the covariant derivative,~\citep[See appendix A of reference][for a detailed explanation]{Masi:2007mn},
i.e.
\begin{equation}
\tau_{\alpha,\beta}= -H_{\alpha\,\beta}\left(\Phi\right)=F_{\alpha,\beta}
-\Gamma^{\gamma}_{\alpha\,\beta}F_{\alpha},
\end{equation}
where, ${\bf F}=-{\bf \nabla} \Phi$. The Christophel symbols, $\Gamma$, and the force ${\bf F}$, must be written in the
selected coordinate system. For our specific case of streams with axial symmetry, it is convenient to use cylindrical
coordinates, defined as in Figure~(\ref{fig:filCoords}); then, in this particular case the tidal force takes the form:

\begin{eqnarray}
{\bf F}_t\left({\bf r}, d{\bf r}\right) &=& \tau_{r\,r}\,d r \, \hat{\bf e}_r + \tau_{\theta\,\theta}\, r\,d\theta
\,\hat{\bf e}_{\theta} \nonumber \\
&=& \dfrac{\partial F\left(r\right)}{\partial r} d r\,\hat{\bf e}_r+ F\left(r\right)d \theta \,\hat{\bf e}_{\theta}.
\label{eq:tidalforce}
\end{eqnarray}

Using the expressions for the force of each of our stream models, eqs. \eqref{eq:stream1D} to \eqref{eq:streamcusp} respectively, it is
straightforward to obtain the expressions for the corresponding tidal forces, given by eq. \eqref{eq:tidalforce}. As an example, the tidal force
in the 1-D stream case is,
\begin{equation}
\mathbf{F}_{t}^{S-1D}=\dfrac{2\,G\,\lambda_{\rm st}}{R^2} dr\,\hat{\mathbf{e}}_r - \dfrac{2\, G\, \lambda_{\rm st}}{R} d\theta
\,\hat{\mathbf{e}}_{\theta},
\end{equation}

Figure~(\ref{fig:tidalforceFils}) shows radial component of the tidal forces, for streams with $\lambda_{\rm st}=1$, or $F_0=1$, accordingly; and
$\gamma=3$ for the core and cusp models is assumed. For the S-1D and S-CD models is the same stretching force outside the stream; once inside the
finite thickness stream, the tidal effect turns into a constant compression.  The Core stream presents a simmilar behaivor as the S-CD, i.e, in some
range gives a compresive force, and in other a streaching one. The Cusp stream gives always a compresive force. The tangential component of the tidal
force coincides with the force, and the effect is always compressive for all the models.

\subsection*{The fractional energy change}

Now, to quantify the effect of an encounter with a stream we will focus on changes to the internal energy of the
target, which is mostly kinetic. One needs to compute the integral  of the tidal force over time to get the kick in the velocity
during to the encounter with a stream, 
\begin{equation}
 \delta \mathbf{v}=\int{\mathbf{F}_t \lp \mathbf{R},\delta \mathbf{r} \rp} \; \rm{d} t,
\label{eq:inteffect}
\end{equation}
where $\mathbf{R}$ is the cylindrical position vector of the target center with respect to the closest stream
point, and $\delta \mathbf{r}$, is the position vector of an element of the target, with respect to the target center.
Both, $\mathbf{R}$ and $\delta \mathbf{r}$ are functions of time. The integral in eq.\eqref{eq:inteffect} should be made
over the time that the encounter takes place, in general this could be accomplished by means of numerical simulations. However,
this problem can be treated analytically in two opposite regimes: If the encounter time is a lot shorter than the internal
dynamical time of the target, then we can take $\delta \mathbf{r}$ as fixed and just take into account the variation of
$\mathbf{R}(t)$, this is the so called, {\it{impulsive regime}}. The opposite regime is when the encounter time is a lot longer
than the internal dynamical time of the target. In this case we can average the effect of the tidal force over an entire dynamical
timescale of the target and then use the average force as the integrand in eq. \eqref{eq:inteffect}, this is
the {\it{adiabatic regime}}.  

The systems of interest to this work are suitable to be worked in the impulsive regime, so we will estimate the effect of a single
encounter using this approximation. The kick in velocity will be obtained by eq.  \eqref{eq:inteffect} once we substitute
the corresponding tidal force, \eqref{eq:tidalforce}, for the different models. But before doing this we must consider the
geometry of the problem. There are two issues here: first, the geometry of the encounter stream-target will define the form of
$\mathbf{R}(t)$; and second, the orientation of the target with respect to the stream will define the
components of $\delta \mathbf{r}$.

\begin{figure}[h!]
\begin{subfigure}{0.5\textwidth}
\centering
\includegraphics[width=0.6\textwidth]{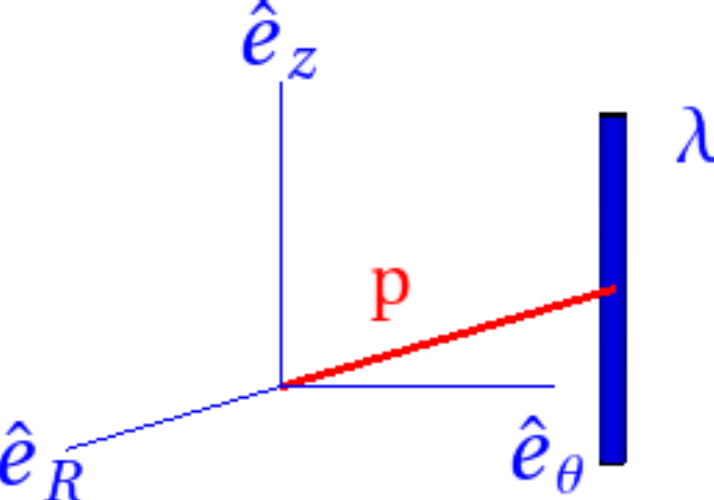}
\caption{}
\label{fig:filCoords}
\end{subfigure}
\begin{subfigure}{0.5\textwidth}
\begin{center}
\includegraphics[width=0.5\linewidth]{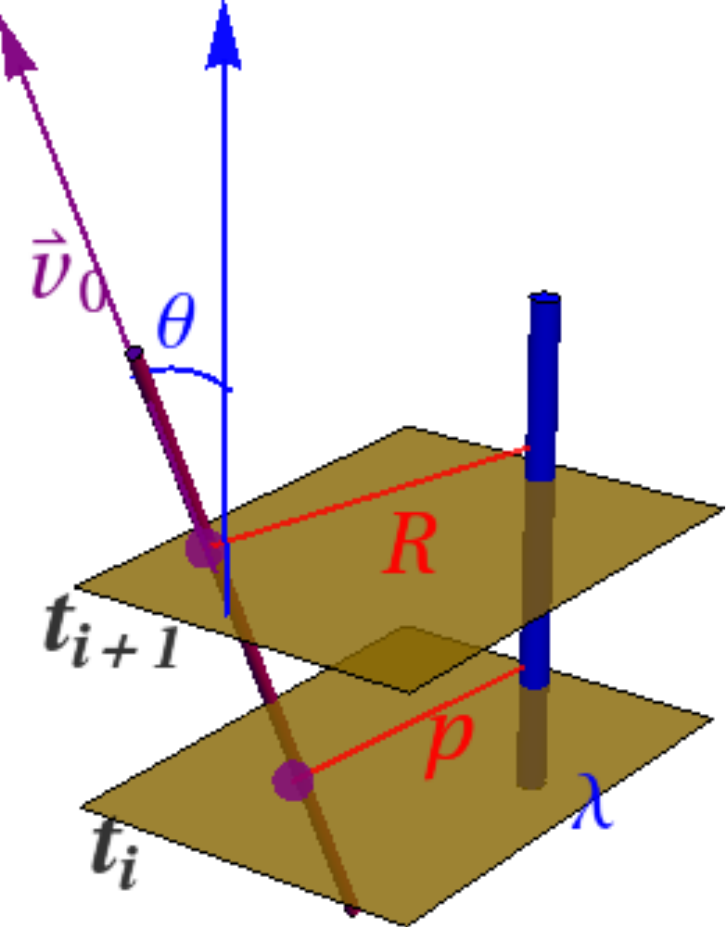}
\caption{}
\label{fig:relativedirection}
\end{center}
\end{subfigure}
\caption{(a)Coordinate frame used for the tidal tensor computation. (b) Relative direction of movement definition.}
\end{figure}

\paragraph{The geometry of the encounter stream-target}

Let us introduce here the concept of the impact parameter, $p$, for encounters with streams. It will be defined as the closest
distance between the stream and the target, with the particularity that it will always be orthogonal to the to the $z$ axis
of the cylinder. Now, lets see the relation of the impact parameter with the vector position $R$ of the target with respect to the
stream. Lets suppose we define the impact parameter at the time $t=t_i$, it should lie in the plane normal to $z$. At this time 
the impact parameter, $p$, and the vector position $R$ coincide. Now as the time evolve, the target will move, in some direction
with respect to the stream, so that the vector position will not coincide any more with the impact parameter. The
difference between them will be just the distance the target, moving at velocity $v_0$ relative to the stream, has
traveled. This distance will depend the relative direction of movement between the stream and target, being $\theta$ the 
angle that characterize this relative direction, as it is shown in Figure~(\ref{fig:relativedirection}). So, the equation that gives the
temporal variation of the target-stream magnitude vector position, $R$, is:

\begin{equation}
 R \lp t; p, \theta, v_0 \rp= \sqrt{p^2+\lp v_0\, t \sin \theta \rp^2}
\end{equation}
One can verify that $\theta=0$ correspond to the case in which the stream and target are moving parallel, at  constant impact parameter. 

\paragraph{The geometry of the target orientation}
Now it is time to turn our attention to the orientation of the target with respect to the stream. What we need to do is to
express the vector position of an element of the target, $\delta {\bf r}$, in the reference frame defined by the stream (Figure \ref{fig:filCoords}),
that is to express $\delta \mathbf{r}$ with respect to the triad of unit vectors $(\hat{e}_R,\hat{e}_{\theta},\hat{e}_z)$.

\begin{figure}[h!]
        \begin{subfigure}{0.33\textwidth}
                \includegraphics[width=\textwidth]{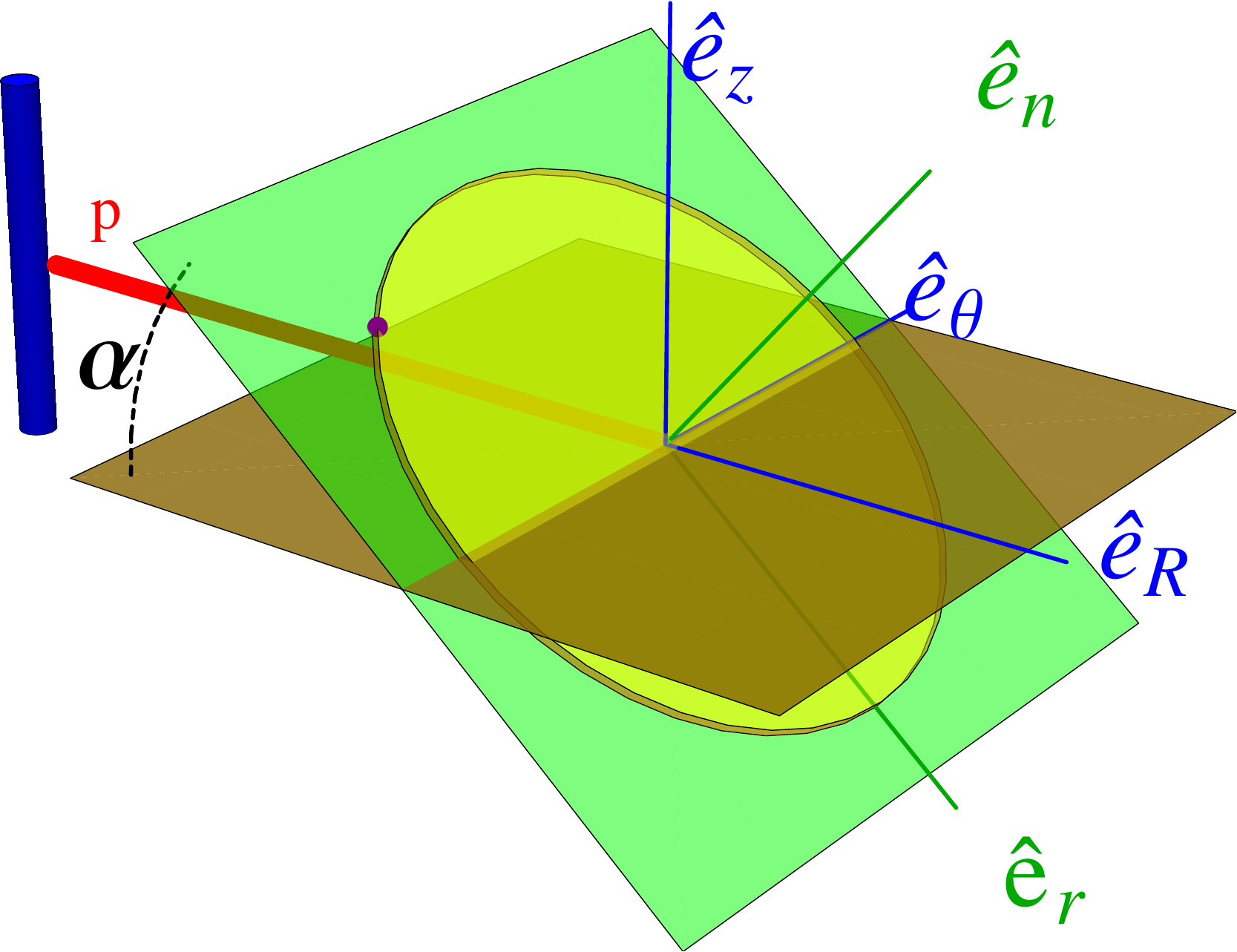}
                \caption{}
                \label{fig:streamsanglesa}
         \end{subfigure}
          \begin{subfigure}{0.33\textwidth}
                \includegraphics[width=\textwidth]{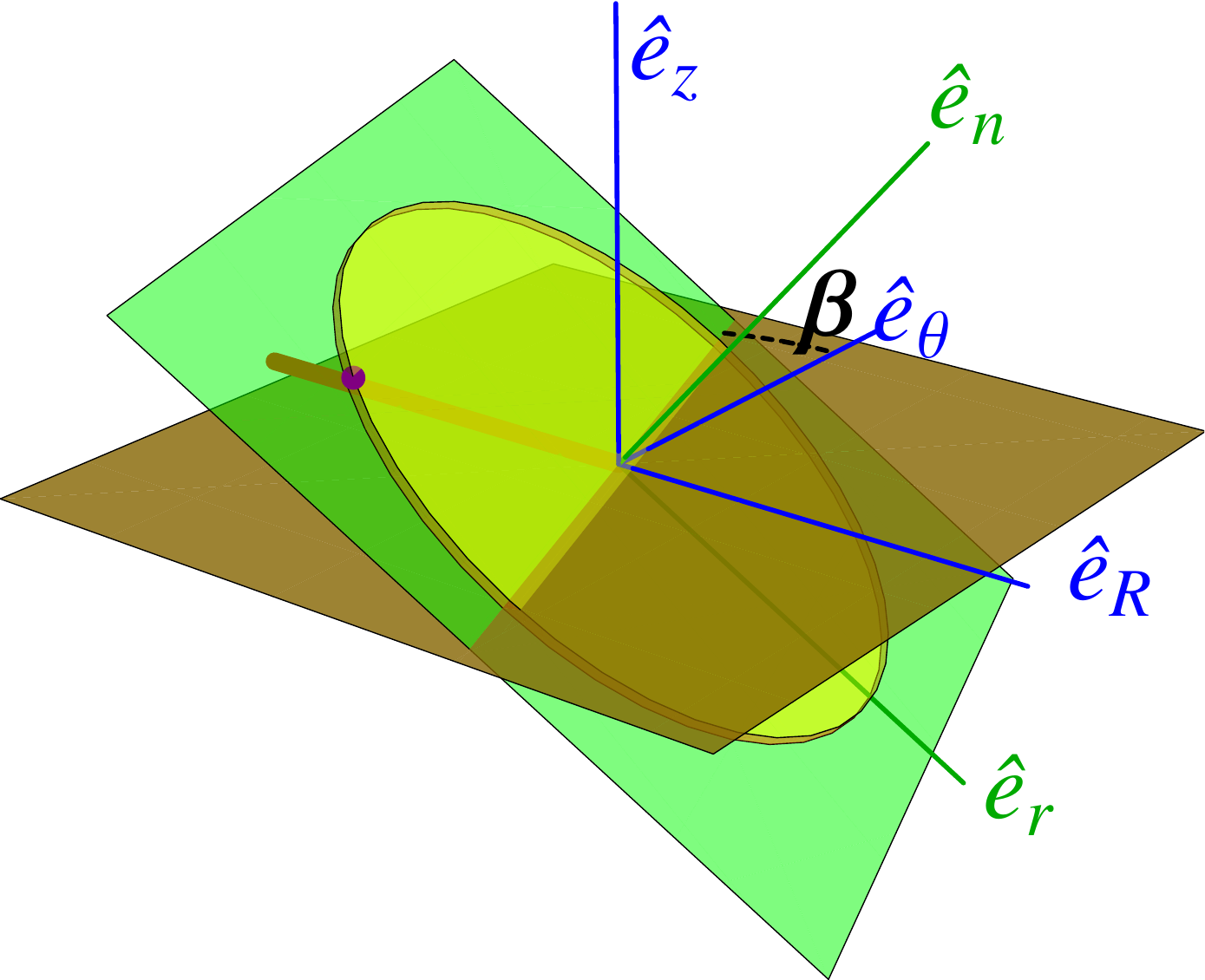}
                \caption{}
                \label{fig:streamsanglesb}
         \end{subfigure}
          \begin{subfigure}{0.33\textwidth}
                \includegraphics[width=\textwidth]{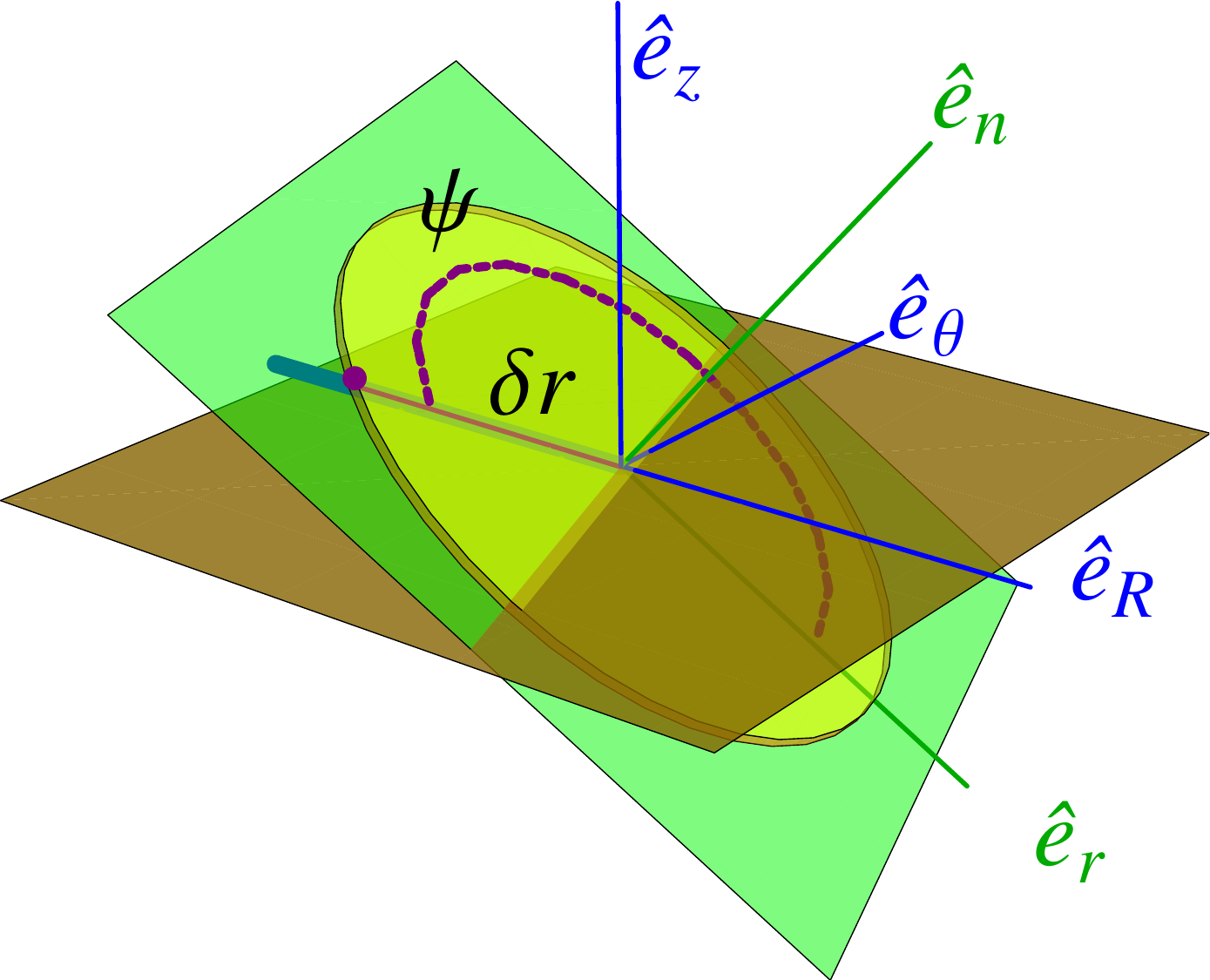}
                \caption{}
                \label{fig:streamsanglesc}
         \end{subfigure}
         \caption{We show the definitions of the rotation angles $\alpha$, $\beta$ and  $\psi$  that allow us to write the
vector position of an element of the target,  $\delta r$ (that lies in the target plane --green--), in the stream reference frame
coordinates (brown plane).}
\label{fig:streamsangles}
\end{figure}

One can describe this transformation by means of a three angle rotation:
\begin{enumerate}
 \item Rotation of the target plane by an angle $\alpha$ with respect to the $\hat{e}_{\theta}$ axis (Figure \ref{fig:streamsanglesa}).
 \item Rotation of the target plane by an angle $\beta$ with respect to the original $\hat{e}_{z}$ axis (Figure \ref{fig:streamsanglesb}).
 \item Rotation on the target plane by an angle $\psi$ with respect to the normal $\hat{e}_n$ of the target plane (Figure \ref{fig:streamsanglesc}).
\end{enumerate}

The first 2 angles defines the spatial orientation of the target plane with respect to the stream plane, the
third angle is just the orbital phase of the target, see Figure \ref{fig:streamsangles} for clarification in the definition of
the angles. It is clear that $-\pi/2 \leq \alpha \leq \pi/2$,  $0 \leq \beta \leq \pi$, and $0\leq\psi\leq 2\,\pi$. The rotation matrix  for the
individual
rotations are: 

\begin{eqnarray}
\begin{array}[4cm]{ccc}
\lambda_{\beta} = \lp \begin{array}{ccc} \cos \lp \beta \rp & \sin\lp \beta \rp & 0 \\
-\sin \lp \beta \rp & \cos \lp \beta \rp & 0 \\ 0 & 0 & 1 \end{array} \rp, \;  & \lambda_{\alpha} = \lp
\begin{array}{ccc} \cos \lp \alpha \rp & 0 & - \sin \lp \alpha \rp   \\
0&1&0 \\ \sin \lp \alpha \rp &0& \cos \lp \alpha \rp \end{array} \rp , \; & \lambda_{\psi} = \lp
\begin{array}{ccc} \cos \lp \psi \rp & \sin \lp \psi \rp & 0 \\ -\sin \lp \psi \rp & \cos \lp \psi \rp & 0 \\ 0 & 0
& 1 \end{array} \rp .
\end{array}
\end{eqnarray}

To get the proper final rotation, we should apply them in the order: first the $\beta$-rotation, followed by the $\alpha$-rotation and finally
the $\psi$-rotation. The inversion in the order of the first two is so that the pivot for the $\alpha$-rotation is with respect to the rotated
$\hat{e}_{\theta}$ axis. If we do it in the inverse order, then the pivot for the second rotation ($\beta$) is the original $\hat{e}_z$ axis, and
not its transformed version, which complicates the algebra. The expression of vector $\delta \mathbf{r}$ using in reference frame attached to stream,
then,  is:
\begin{eqnarray}
\delta \mathbf{r} &=& \delta r \left\lbrace \left[\cos \lp \alpha \rp \, \cos \lp \beta \rp \cos \lp \psi \rp -\sin \lp
\beta \rp \, \sin \lp \psi \rp \right] \hat{\mathbf{e}}_R \right. \nonumber \\
&+& \left. \left[\cos \lp \alpha \rp \, \cos \lp \beta \rp \sin \lp \psi \rp +
\cos \lp \beta \rp \, \sin \lp \psi \rp \right] \, \hat{\mathbf{e}}_{\theta} 
- \left[ \cos \lp \psi \rp \sin \lp \alpha \rp \right] \hat{\mathbf{e}}_z\right\rbrace 
\label{eq:deltrtransformed}
\end{eqnarray}

Finally, the perturbation energy (per unit mass) transfered to the internal dynamics of an extended target system, by one single stream
encounter, is simply

\begin{equation}
\Delta E=\Delta (1/2 v^2) = (\mathbf{v} \cdot \Delta \mathbf{v}) + 1/2 (\Delta \mathbf{v})^2 \sim 1/2 (\Delta v)^2                
\end{equation}

where in the last step, we have neglected the linear term between the velocity of the target planet, $\mathbf{v}$, and the tidal velocity kick,
$\Delta \mathbf{v}$ (random orientation). We now compare this perturbation energy with the binding energy of the target: $E_b=-G\,M_{\rm c}/(2\,a)$,
and
substitute equations \eqref{eq:deltrtransformed} and \eqref{eq:tidalforce} in eq.\eqref{eq:inteffect}. The integration was done considering $\delta
{\bf r}$ constant, i.e. $\delta r, \alpha, \beta$ and $\psi$ constants during the encounter. The results are the following, for the different stream
models.

\begin{subnumcases}{\dfrac{\langle\Delta E\rangle_{\rm st}}{|E_b|}=\dfrac{4\,G\, \pi^2 \lambda^2}{M_{c}\, v^2_{0}\, {\sin\lp
\theta\rp}^2\,p^2 } a^3}
 \varmathbb{T}\lp\alpha, \psi\rp & (1D)\\
 \varmathbb{T}\lp\alpha, \psi\rp \,{\varmathbb{B}^2\lp R_0/p\rp} &  (CD)\\
  \varmathbb{C}\lp R_0/p,\alpha,\beta,\psi\rp  & (Core) \\ 
 \varmathbb{D}\lp R_0/p,\alpha,\beta,\psi\rp & (Cusp).
\end{subnumcases}

where the geometric functions, $\varmathbb{T}$, $\varmathbb{B}$, $\varmathbb{C}$, and $\varmathbb{D}$ are given by:

\begin{eqnarray}
\varmathbb{T}&=& \left[\cos^2\lp\alpha\rp\, \cos^2\lp\psi\rp+
\sin^2\lp\psi\rp\right]  \\
\varmathbb{B}&=& \frac{2}{\pi}\left[ \arctan \lp\frac{R_0/p}{\sqrt{1-R_0/p^2}}\rp-R_0/p \sqrt{1-R_0/p^2}\right]
 \qquad R_0/p \leq 1  \\
&=& 1, \qquad \qquad \qquad  \qquad \quad \qquad \qquad \qquad R_0/p \geq 1 \nonumber \\
\varmathbb{C}&=& {I_1}^2(R_0/p)\left[\cos\lp\alpha\rp\,\cos \lp \beta \rp\,
\cos\lp\psi\rp-\sin\lp\beta\rp\,\sin\lp\psi\rp\right]^2 + \\
& & {I_2}^2 (R_0/p)\left[\cos\lp\alpha\rp \,\cos \lp \psi \rp\,\sin \lp \beta \rp +
\cos\lp\beta\rp\sin\lp\psi\rp\right]^2, \nonumber \\ 
& &\enskip I_1= 
\frac{2}{\pi}\int_{0}^{\infty}{\frac{\sqrt{1-\tau^2}-R_0/p}{\left[\sqrt{1+\tau^2}+R_0/p \right]^3} d \tau}
\nonumber \\
& &\enskip I_2= \frac{2}{\pi}\int_{0}^{\infty}{\frac{1}{\left[\sqrt{1+\tau^2}+R_0/p \right]^2} d \tau} \nonumber \\
\varmathbb{D}& =& {I_3}^2(R_0/p)\left[\cos\lp\alpha\rp\,\cos\lp\beta\rp\,
\cos\lp\psi\rp-\sin\lp\beta\rp\,\sin\lp\psi\rp\right]^2 + \\
& & {I_4}^2 (R_0/p)\left[\cos\lp\alpha\rp \,\cos\lp\psi\rp\,\sin\lp\beta\rp +
\cos\lp\beta\rp\sin\lp\psi\rp\right]^2, \nonumber \\
& &\enskip I_3= 
\frac{2}{\pi}\int_{0}^{\infty}{\frac{d \tau}{\left[R_0/p\,\sqrt{1+\tau^2}\right]^2} d \tau} \nonumber \\
& &\enskip I_4= 
\frac{2}{\pi}\int_{0}^{\infty}{\frac{d \tau}{\sqrt{1+\tau^2}\left[R_0/p +\sqrt{1+\tau^2}\right]} d \tau} \nonumber
\end{eqnarray}

\begin{figure}[t]
        \begin{subfigure}{0.5\textwidth}
                \centering
                \includegraphics[width=\textwidth]{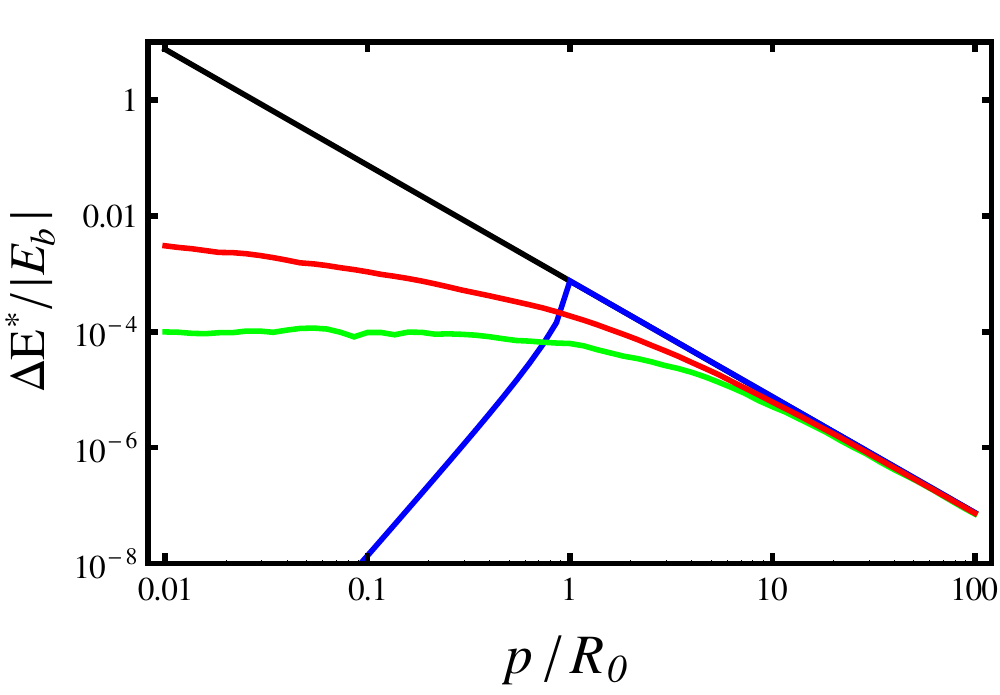}
                \caption{Perturbation energy produced by the streams, as a function of the impact parameter, $p$. The models are the same as
in Figure \ref{fig:forceFils}}
                \label{fig:deltaEstream}
        \end{subfigure}%
        \quad
        \begin{subfigure}{0.5\textwidth}
                \centering
                \includegraphics[width=\textwidth]{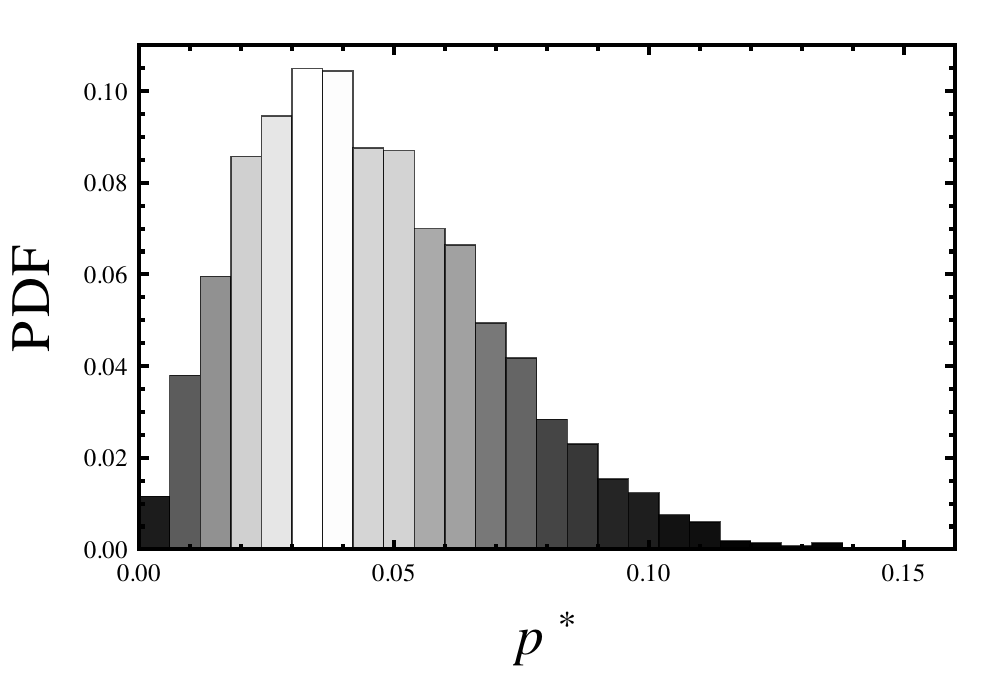}
                \caption{Impact parameter probability distribution functions for dark streams calculated by the sampling procedure described in the
text.}
                \label{fig:PDF}
        \end{subfigure}
        \caption{}
\end{figure}

All the functions $\Delta E/E_b$ for the different models, converge to that of the S-1D model for large impact parameters, see Figure
\ref{fig:deltaEstream}. 

\subsection{Impact parameters for streams.}
\label{sec:appendix4}
 In order to generate the distribution function of impact parameters for the streams, we follow the next procedure. We start by tossing a given
number of random lines within a region of the 3D-space. Then we also tose a collection of uniformly distributed points, within the same region, and
determine the distance from each point to the closest line, that is the impact parameter. This results in a sampling of the underlying distribution of
impact parameters that is completely independent of the introduction of a sample volume. We show the resultant probability distribution in
Figure~(\ref{fig:PDF}). We still need to relate this realization to a specific distribution of matter with a certain spatial mass density, $\rho$, and
linear mass density for the lines, $\lambda$, so that the impact parameter denoted by $p^{*}$ can have the correct dimensions according to the
specific physical situation. The way to do this to introduce a sample volume to measure the
mass density around each of the tests points. We did several experiments, keeping fixed the total number of lines in the
realization, as well as the realization volume, but changing the size of the sample volume, $R^{*}$, and we found that the number
density of streams varies with it, but the mass density doe not. Thus, in our procedure, the mass density is a well defined
physical quantity whose value is independent of the size of the region used to compute it.
In the practice, we normalized the linear mass density of the streams to the unity so that the mass density we get from the
realization is $\rho^{*}$. By doing a dimensional analysis we get to the conclusion that the relation between the impact
parameter, $p^{*}$, and the physical impact parameter $p$, is given by: $p=\lp\sqrt{\lambda_{\rm st}\,\rho^{*}/\rm f_{\rm s}\,\rho}\rp\,p^{*}$, where
$\lambda_{\rm st}$ and $\rho$ are given in physical units, and $\rho^{*}$ is the numerical mass density obtained in a given realization.\\
We will use the impact parameter distribution function shown in Figure \ref{fig:PDF}, that corresponds to a numerical mass density of
$\rho^{*}=1205.08$; the $\lambda_{\rm st}$, $\rm f_{\rm s}$, and $\rho$ parameters will be fixed according to the specific situation under study.

\end{document}